\documentclass{aa}  

\usepackage{graphicx,natbib}
\usepackage{multirow}

\usepackage{txfonts}
\usepackage[breaklinks=true,colorlinks=blue,urlcolor=blue,citecolor=blue,linkcolor=blue]{hyperref}

\newcommand{\asecdot}[2]{\mbox{#1$\stackrel {\prime \prime}{_{\bf \cdot}}$#2}}
\newcommand{\RHK}{$R^\prime_\mathrm{HK}$}

\newcommand{ \DSmod}{$\Delta S/S_{mod}$}

\newcommand{\Ms}{M$_\odot$}
\newcommand{\Snu}{$S_\mathrm{obs}(\nu)$}

\newcommand{\Smod}{$S_\mathrm{mod}(\nu)$}

\newcommand{\teff}{$T_\mathrm{eff}$}
\newcommand{\Tb}{$T_\mathrm{B}$}
\newcommand{\Tbp}{$T_\mathrm{B}(\nu)$}
\newcommand{\Ro}{$R_\mathrm{o}$}

\begin{document}

    \title{
    EMISSA - Exploring Millimeter Indicators of Solar-Stellar Activity} 
   \subtitle{I. The initial millimeter -- centimeter main-sequence star sample}
    \titlerunning{EMISSA I. Initial main-sequence star sample}

    
   \author{A. Mohan\inst{1,2}
         , S. Wedemeyer\inst{1,2}
         , S. Pandit\inst{1,2}
          \and 
          M. Saberi\inst{1,2}
          \and 
          P. H. Hauschildt\inst{3} 
          }
    \institute{Rosseland Centre for Solar Physics, University of Oslo, Postboks 1029 Blindern, N-0315 Oslo, Norway\\
    \and
    Institute of Theoretical Astrophysics, University of Oslo, Postboks 1029 Blindern, N-0315 Oslo, Norway\\
    \email{atulm@astro.uio.no}
    \and
    Hamburger Sternwarte, Gojenbergsweg 112, 21029 Hamburg, Germany
}

             
\authorrunning{Mohan, A. et al.}

   \date{Received  - ; accepted  - }

 
  \abstract
  {
  %
Due to their wide wavelength coverage across the millimeter to centimeter (mm -- cm) range and their increased sensitivity, modern interferometric arrays facilitate observations of the thermal and non-thermal radiation that is emitted from different layers in the outer atmospheres of stars.    }
    {
     We study the spectral energy distribution (\Snu) of main-sequence stars based on 
    archival observations in the mm -- cm range with the aim to study their atmospheric stratification as a function of stellar type.   } 
    %
  {
    The main-sequence stars with significant detection in mm bands were identified in the ALMA Science Archive. 
    These data were then complemented with spectral flux data in the extreme ultraviolet (EUV) to cm range as compiled from various catalogues and observatory archives. 
    We compared the resultant \Snu\ of each star with a photospheric emission model (\Smod) calculated with the PHOENIX code.  
    The departures of \Snu\ from \Smod\ were quantified 
    in terms of a spectral flux excess parameter (\DSmod) and studied  as a function of stellar type.} 
  {
  The initial sample consists of 12~main-sequence stars across a broad range of spectral types from A1 to M3.5 and the Sun-as-a-star as reference. 
    The stars with \teff\,$ = 3000 - 7000$\,K (F -- M type) showed a systematically higher \Snu\ than \Smod\ in the mm -- cm range. Their \DSmod\ exhibits a monotonic rise with decreasing frequency. The steepness of this rise is higher for cooler stars in the \teff\,= 3000 -- 7000\,K range, although the single fully convective star (\teff\ $\sim$ 3000\,K) in the sample deviates from this trend. 
    Meanwhile, \Snu\ of the A-type stars agrees with \Smod\ within errors. 
  }
  {
  The systematically high \DSmod\ in F -- M stars indicates hotter upper atmospheric layers, that is, a chromosphere and corona in these stars, like for the Sun. The mm -- cm \DSmod\ spectrum offers a way to estimate the efficiency of the heating mechanisms across various outer atmospheric layers in main-sequence stars, and thereby to understand their structure and activity. 
  We emphasise the need for dedicated surveys of main-sequence stars in the mm -- cm range.}


    

   \maketitle
%

\section{Introduction}
The  millimeter to centimeter (mm--cm)  waveband ($\approx 10 -- 1000$\,GHz) offers a unique but so far not systematically used way to explore the atmospheric structure of main-sequence stars. As the activity of a star is closely linked to the physical state and dynamics of the plasma in its  atmospheric layers  \citep[e.g.][]{Noyes84_RHK,Pace13_RHK_Drastic_variability}, the mm--cm waveband thus provides complementary means of assessing stellar activity. 
Because it is so far the only spatially resolved main-sequence star with a large wealth of data and because it consequently is the star we understand best theoretically, the Sun is a good template star that provides insight into the most relevant emission mechanisms in the mm -- cm band. The results for the Sun may be transferred to other Sun-like  stars of spectral type (FGK type) \citep{2004NewAR..48.1319W} with possible implications for a wider range along the main sequence. 
In the Sun, the mm continuum ($\approx$ 30 -- 1000\,GHz) is primarily produced by thermal free-free emission,  
whereas non-thermal contributions are only expected to become significant under extreme conditions like strong magnetic field enhancements and flares.
The height of the layer above the photosphere from which the continuum emission originates increases  with decreasing observing frequency \citep[see][and references therein for solar mm radiation]{2016SSRv..200....1W}.
Like for the Sun, the emission in the mm--cm waveband can be expected to originate primarily from different (optically thick) atmospheric layers at different heights in the chromosphere to transition region in Sun-like stars (FGK type).
A recent study of $\alpha$~Cen\,A\&B (G2V \& K1V) across 17 -- 600\,GHz revealed a 
brightness temperature \Tbp\ for both stars that decreased with frequency and is thus consistent with a chromospheric temperature rise as  expected for these spectral types \citep{2018MNRAS.481..217T}.

For high-mass main-sequence stars (OBA type), the outer atmospheric structure is expected to be quite different from that of Sun-like stars. It is expected that the average temperature rise as seen in the chromosphere and corona of Sun-like stars is absent in high-mass stars.
This is primarily because unlike  Sun-like stars, OBA stars are not expected to have an outer convection zone that generates and drives dynamic phenomena
that play a vital role in generating the hot outer atmospheric layers in Sun-like stars \cite[see e.g.][ for a review]{Donati09_Rev_Bfield}.
Observations of $\alpha$\,CMa (A1V) \citep{2019ApJ...875...55W} and $\alpha$\,PsA  \citep{2016ApJ...818...45S} (A4V) show indeed no significant chromospheric emission. 
Rather, the mm flux as observed with the Atacama Large Millimeter/submillimeter Array (ALMA) 
seems to be consistent with purely photospheric emission 
 \citep[for $\alpha$~CMa, see][]{2019ApJ...875...55W}.

Using the Rayleigh-Jeans approximation, the spectral flux density, $S_\nu$, can be converted into the brightness temperature \Tb, as described by
\begin{equation}
S_\nu = \frac{2\ k\ T_B\ \nu^2}{c^2}\ \Omega_s \quad,
\end{equation}
where $k$ is the Boltzmann constant and $\Omega_s$ is the solid angle subtended by the star. The latter is given as $\Omega_s = \pi R_*^2/D^2$, where $R_*$ is the radius of the star and $D$ is the distance to it. 
Because the emission primarily originates from optically thick layers in the mm range, the brightness temperature spectrum, \Tbp, is closely related to the temperature in the continuum-forming layer at a height $z\,(\nu)$ in the atmosphere. 
Consequently, the observable brightness temperature spectrum ($T_B\,(\nu)$) is a proxy to the atmospheric temperature stratification, $T\,(z\,(\nu))$ 
\citep[see e.g. the semi-empirical models for the Sun by][]{1981ApJS...45..635V}. 
Systematic mm -- cm observations of main-sequence stars of different spectral types and ages 
will therefore provide important constraints for the evolution of stellar atmospheric structure and activity. 
Comparing these observations with magnetic field properties will help to better understand the connection between atmospheric structure, activity, and magnetic field as predicted by various models \citep[e.g.][]{1996ApJ...457..340K,2018ApJ...862...90G} and inferred from multi-parameter studies of stars in different surveys \citep[e.g.][]{Donati09_Rev_Bfield,Vidotto14_B_Vs_age_n_rot, Lund20_stelar_helicity,2020AJ....160..219F}.

However, only a few  main-sequence stars have been detected in the mm -- cm range so far compared to the wealth of detections in the IR to X-ray range. 
The low detection rate so far is a direct consequence of the low 
expected flux densities of even the nearest Sun-like stars, which are lower than $\sim 100$\,$\mathrm{\mu}$Jy (1\,Jy $= 10^{-26}\,$Wm$^{-2}$Hz$^{-1}$) in mm bands and well below 100\,$\mathrm{\mu}$Jy at cm wavelengths \citep{2004NewAR..48.1319W}. A reliable detection at a single frequency therefore requires an instrument capable of providing a sensitivity of about a few tens of $\mu$Jy within a reasonable integration time of a few hours at most and over a spectral bandwidth of typically $\sim$10\,GHz. 
%
In addition, a significant fraction of main-sequence stars belong to binary systems with separations that can be as small as a few arcseconds and/or have debris disks with an inner edge close to the star \citep[see e.g.][]{2014ApJ...788L..37G,2007ApJ...655L.109M,2007ApJ...658..584L}. 
An example of the former case is the $\alpha$\,Cen\,A\&B 
system at a distance of 1.3\,pc with an angular separation of just 4$^{\prime\prime}$ \citep{2016A&A...594A.109L}. 
$\epsilon$\,Eri
at a distance of 3.2\,pc is an example of a star with a debris disk that has an inner edge at $\sim10^{\prime\prime}$ from the star. 
Consequently, in order to measure the stellar flux without contamination from a companion or a disk, the angular resolution of an observation must be high enough to separate the star from these potential components. 
The required angular resolution for observations in the mm -- cm bands warrants the use of interferometric arrays. 
Because it is the most sensitive telescope with the largest number of baselines and a large collecting area in mm bands, ALMA\footnote{\href{https://almascience.nrao.edu/about-alma/alma-basics}{https://almascience.nrao.edu/about-alma/alma-basics}} 
is currently the only telescope that meets the requirements in the mm range.
In the cm bands, the Australia Telescope Compact Array\footnote{\href{https://www.narrabri.atnf.csiro.au/observing/users\_guide/html/atug.html}{https://www.narrabri.atnf.csiro.au/observing/users\_guide/}} (ATCA) and the Expanded Very Large Array (JVLA, \citealp{Perley11_EVLA}) are examples of new-generation arrays that can provide such data. 

Most of the studies of main-sequence stars in radio wavebands used frequencies below 10\,GHz so far, much of which have been on active M dwarfs \citep[see,][for a review]{2002ARA&A..40..217G}. 
This is because the flares, especially in active M dwarfs and Sun-like stars, cause coherent radio emissions brighter by several orders of magnitude than the background free-free emission at low radio frequencies \cite[e.g.][]{ 1994SSRv...68..261B,Lim96_Proxima_ATCA, 2018ApJ...856...39C, 2019ApJ...871..214V, 2020MNRAS.494.4848D, Zic20_typeIV_ProximaCen}. At high frequencies, the flaring rates are low and the emission \Tb\ is also much lower, requiring sensitivity limits that are hard to achieve with earlier-generation arrays even for the nearby stars \citep{1990SoPh..130..265B, 2004NewAR..48.1319W}.  
However, a few M dwarfs were detected close to 10\,GHz by the VLA, namely UV\,Cet and EV\,Lac \citep{2005ApJ...621..398O, 2006astro.ph..9389G}. 
More recently, the quiescent and flare emission from the nearby M dwarf Proxima Cen was studied with ALMA \citep[e.g.][]{AngladaEscude12,MacGregor18_proxima_flares,2021ApJ...911L..25M}.
\cite{Villadsen14_First_detect_SLS_inRadio_VLA} presented the first detection of thermal emission in the cm band using JVLA from Sun-like stars, $\tau$\,Cet, $\eta$\,Cas\,A, and 40\,Eri\,A.
$\tau$\,Cet was also observed with ALMA by \citet{2016ApJ...828..113M}, but the emission was partly contaminated by the edge-on disk emission.
Using high-resolution ALMA observations, \citet{2017MNRAS.469.3200B} were able to distinguish the emission from the K-dwarf $\epsilon$Eri from its debris disk unlike the IR-band observations of the star \citep{2015A&A...576A..72L, ChavezDagostino16}.
\cite{Bastian18_EpsEri}, \cite{2019ApJ...871..172R}, and \cite{suresh20_EpsEri_RadioSEDmodel} extended this up to 2\,GHz using the EVLA.
The closest binary system $\alpha$\,Cen\,A\&B is another repeatedly observed pair of stars that has been observed across all available bands of ALMA, except band 10. The observations were carried out during two different observing cycles separated by 2\,yr \citep{2013A&A...549L...7L, 2015A&A...573L...4L, 2016A&A...594A.109L,2019arXiv190403043L}.
A recent study by \cite{2021AJ....162...14A} at $\approx$\,340\,GHz reported similar flux for the binary stars as \cite{2016A&A...594A.109L}.
\cite{2018MNRAS.481..217T} modelled the mm -- cm spectrum of $\alpha$\,Cen\,A\&B using various scaled solar atmospheric models and demonstrated the existence of a solar-like chromosphere that becomes hotter with increasing distance from the surface. The authors also added 17\,GHz flux data points from ATCA to the existing spectra. 
Following a similar modelling approach as \cite{White20_MESAS}, they modelled their observations of F-type stars, $\gamma$\,Lep and $\gamma$\,Vir\,A\&\,B, at $\approx$ 233 and 344\,GHz.
Other F dwarfs that have reliable detection in the mm band are 61\,Vir \citep{2017MNRAS.469.3518M} and $\eta$Crv \citep{2017MNRAS.465.2595M}.
Of the more massive stars, the closest A-type stars, $\alpha$\,CMa and $\alpha$\,PSa, are the most frequently studied. The former was studied across the $\approx$30 -- 210\,GHz range and the spectrum was modelled by \cite{2019ApJ...875...55W}. The latter garnered more interest primarily due to its circumstellar disk, which was well resolved with ALMA \citep[e.g.][]{2016ApJ...818...45S, 2017MNRAS.466.4201W,  2017ApJ...842....8M, 2017ApJ...842....9M}.

We present an initial version of a database of mm -- cm fluxes for main-sequence stars detected by ALMA  supplemented with other observations from the extreme ultraviolet (EUV) to the radio band.  
The Sun is included as a fundamental reference. 
The resulting observed spectral energy distribution (SED) for each star is then compared to the SED  calculated with the PHOENIX model \citep{1999JCoAM.109...41H} to reveal possible chromospheric contributions. 
The method and initial data set are described in Sect.~\ref{sec:method}. The resulting SEDs and and \Tbp\ are presented in Sect.~\ref{sec:analysis}. Section~\ref{sec:discussion} presents the inferences from a comparison of the model and observed SEDs for stars of different types, followed by our conclusions and outlook in Sect.~\ref{sec:conclus}.

\section{Method and data sets} 
\label{sec:method}
\begin{figure*}[!t]
    \begin{center}
    \includegraphics[width=0.9\textwidth,height=0.45\textwidth]{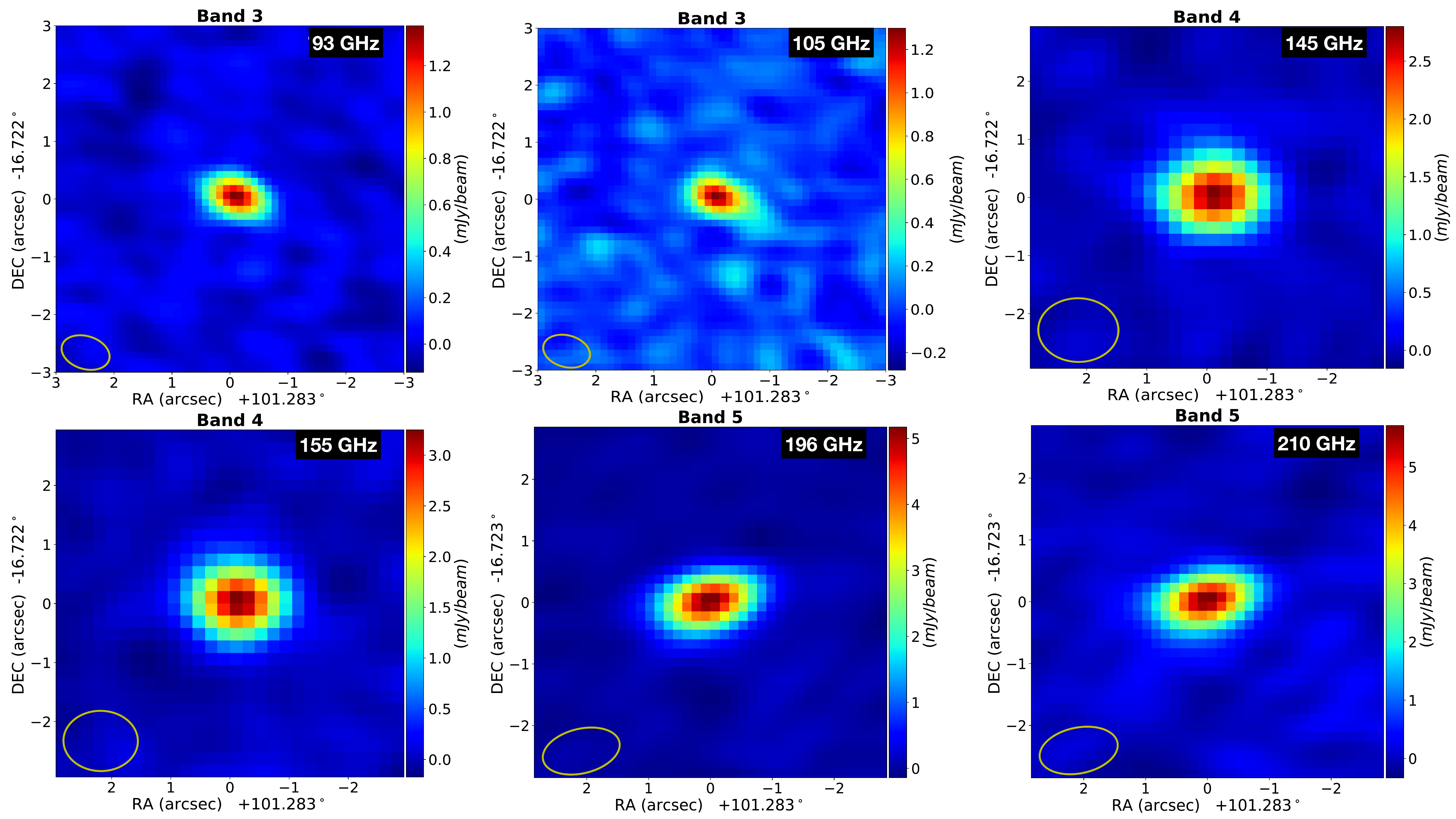}
    \caption{Images of $\alpha$~CMa in various ALMA bands provided by the JVO server. The data were spectrally averaged over a bandwidth of 2\,GHz for all images. The ellipses represent the respective synthetic beams of the individual observations. 
    }
    \label{fig:Sirius_imgs}
    \end{center}
\end{figure*}
%
\begin{sidewaystable}
\begin{tabular}{|l|l|l|l|l|l|l|l|l|l|l|l|}
\hline
\multirow{2}{*}{\textbf{ID}} & \multirow{2}{*}{\textbf{HD/GJ ID}} & \multirow{2}{*}{\textbf{SIMBAD ID}} & \multirow{2}{*}{\textbf{Sp.Type}} & \textbf{RA} & \textbf{DEC} & \textbf{Distance} & \textbf{Mass} & \textbf{Radius} & \textbf{T$_\mathrm{eff}$} & \textbf{Log g} & 
\multirow{2}{*}{\textbf{[Fe/H]}} \\ \cline{5-11} 
 &  &  &  & deg & deg & pc & M$_\odot$ & R$_\odot$ & K & cm/s$^2$ & 
 \\ \hline
A & HD 48915A & Alf CMa & A1V & 101.287[3] & -16.716 [3] & 2.64$\pm$0.01 [14] & 2.15 [4] & 1.71$\pm$0.01 [43] & 9880$\pm$200 & 4.31 & 0.29$\pm$0.5 [3] 
\\ \hline
B & HD 216956 & Alf PsA & A4V [33] & 344.413[3] & -29.622 [3] & 7.70$\pm$0.03 [14] & 1.97 [4] & 1.84$\pm$0.02 [36] & 8900$\pm$300 & 4.20 & 0.45$\pm$0.2 [3] 
\\ \hline
C & HD 110379 & Gam Vir A & F0V [29] & 190.4151 [3] & -1.44939 [3] & 11.62$\pm$0.09 [14] & 1.57 [60] & 1.39$\pm$0.19 [*] & 7100$\pm$100 & 4.35 & -0.07$\pm$0.08 [3] 
\\ \hline
D & HD 110380 & Gam Vir B & F0V [29] & 190.4148 [3] & -1.44940 [3] & 11.62$\pm$0.09 [14] & 1.43 [60] & 1.49$\pm$0.2 [*] & 6960$\pm$146 & 4.25 & -0.09$\pm$0.06 [62] 
\\ \hline
E & HD 109085 & Eta Crv & F2V & 188.018 [3] & -16.196[3] & 18.28$\pm$0.06 [14] & 1.48 [12] & 1.52$\pm$0.06 [12] & 6871$\pm$48 & 4.25 & 0.00$\pm$0.06 [51] 
\\ \hline
F & HD 38393 & Gam Lep & F6V & 86.116 [13] & -22.448[13] & 8.88$\pm$0.02 [14] & 1.22 [12] & 1.30$\pm$0.06 [12] & 6599$\pm$556 & 4.30 & -0.02$\pm$0.1 [49] 
\\ \hline
G & HD 128620 & Alf Cen A & G2V & 219.902 [1] & -60.834 [1] & 1.325$\pm$0.007 [38] & 1.11 [16] & 1.25$\pm$0.03 [17] & 5745$\pm$80 & 4.29 & 0.2$\pm$0.02 [50] 
\\ \hline
H & HD 115617 & 61 Vir & G6.5V & 199.601 [3] & -18.311 [3] & 8.53$\pm$0.02 [14] & 0.92 [25] & 0.98$\pm$0.04 [10] & 5618$\pm$70 & 4.42 & 0.02$\pm$0.46 [49] 
\\ \hline
I & HD128621 & Alf Cen B & K1V & 219.896[13] & -60.838 [13] & 1.325$\pm$0.007 [38] & 0.90 [25] & 0.86$\pm$0.02 [17] & 5145$\pm$80 & 4.53 & 0.23$\pm$0.03 [50] 
\\ \hline
J & HD 22049 & Eps Eri & K2V & 53.233 [2] & -9.458 [2] & 3.202$\pm$0.005 [38] & 0.86 [21] & 0.77$\pm$0.02 [21] & 5100$\pm$70 & 4.60 & -0.05$\pm$0.46 [49] 
\\ \hline
K & GJ 2006 A & GJ 2006 A & M3.5V & 6.959[3] & -32.552 [3] & 34.89$\pm$0.09 [14] & 0.55 [12] & 0.56$\pm$0.02 [12] & 3150$\pm$18 & 4.69 & -0.5$\pm$0.1 [3] 
\\ \hline
L & GJ 551 & Proxima Cen & M5.5V & 217.43[3] & -62.68 [3] & 1.3$\pm$0.003 [14] & 0.12 [64] & 0.15$\pm$0.004 [64] & 2990$\pm$44 & 5.16 & -0.07$\pm$0.14 [3] 
\\ \hline
S & Sun &  & G2V &  &  & 4.8E-06 & 1 & 1 & 5772 [21] & 4.44 & \begin{tabular}[c]{@{}l@{}}log(Fe/H) = \\   -4.49 [54]\end{tabular} 
\\ \hline
\end{tabular}%
\vspace{0.12cm}
\caption{Physical properties of ALMA-EMISSA main-sequence stars. Columns 1 to 13 provide the assigned ID, HD/GJ ID, SIMBAD ID, spectral type, RA, DEC, distance, mass, radius, effective temperature, surface gravity, and metallicity of each star with references numbered inside []. Spectral types are obtained from SIMBAD, unless specified and T$_\mathrm{eff}$s from \cite{Soubiran16_Ref3}. $\log$\,g is computed using mass and radius.}
\label{tab:prop}
\end{sidewaystable}

\begin{sidewaystable}
\begin{tabular}{|l|l|l|l|l|l|l|l|l|l|l|l|}
\hline
\multirow{2}{*}{\textbf{ID}} & \multirow{2}{*}{\textbf{SIMBAD ID}} & \textbf{Age} & \textbf{B-V} & \textbf{Vsini} & \textbf{L$_\mathrm{bol}$} & \multirow{2}{*}{\textbf{Log R$^\prime_\mathrm{HK}$}} & \textbf{Log L$_\mathrm{X}$} & \multirow{2}{*}{\textbf{Log R$_\mathrm{X}$}} & \textbf{|B$_\mathrm{lon}$|} & \textbf{Period} & \textbf{R$_\mathrm{o}$} \\ \cline{3-6} \cline{8-8} \cline{10-12} 
 &  & Myr &  & km/s & L$_\odot$ &  & erg/s &  & G & d &  \\ \hline
A & Alf CMa & 242 [43] & 0.0 [5] & 16 [45] & 25.4 [44] & – & 28.82 [58] & – & – & 5.37[*] & – \\ \hline
B & Alf PsA & 400 [46] & 0.094 [5] & 93 [37] & 16.63 [36] & – & – & – & – & 1 [*] & – \\ \hline
C & Gam Vir A & 1140 [23] & 0.36 [34] & 36 [60] & 4.37 [60] & -4.4789 [50] & 29.03 [63] & -5.21 [63] & 15.5$\pm$10.2 [20] & 2.04 [*] & – \\ \hline
D & Gam Vir B & 1140 [23] & 0.36 [59] & 23 [60] & 4.63 [60] & -4.2984 [50] & 29.03 [63] & -5.24 [63] & 21$\pm$16 [20] & 3.19 [*] & – \\ \hline
E & Eta Crv & 1012 [22] & 0.38 [24] & 68 [39] & 4.679 [48] & -4.959 [23] & 28.68 [18] & -5.62 [23] & – & 1.1 [*] & – \\ \hline
F & Gam Lep & 1300 [57] & 0.5 [24] & 10 [40] & 2.291 [41] & -4.77 [11] & 28.3 [13] & -5.71 [23] & – & 6 [41] & – \\ \hline
G & Alf Cen A & 4850 [30] & 0.71 [16] & 2.7 [31] & 1.519 [30] & -5.059 [16] & 27.7 [13] & -6.96 [*] & 28.7 [*] & 22 [31] & 2.223 [*] \\ \hline
H & 61 Vir & 6300 [47] & 0.709 [16] & 3.9 [27] & 0.821 [42] & -5.04 [16] & 26.87 [13] & -6.9 [10] & 30.6 [*] & 12.8 [27] & 2.186 [*] \\ \hline
I & Alf Cen B & 4850 [30] & 0.85 [26] & 1.1 [31] & 0.5 [30] & -4.94 [33] & 27.06 [10] & -6.61 [*] & 51.8 [*] & 36 [32] & 1.989 [*] \\ \hline
J & Eps Eri & 400 [15] & 0.881 [2] & 2.4 [57] & 0.381 [14] & -4.51 [2] & 28.32 [8] & -4.78 [8] & 8.3$\pm$4.9 [20] & 11.8 [21] & 0.366 [8] \\ \hline
K & GJ 2006 A & 6 [35] & 1.5 [6] & 6.2 [7] & 0.046 [14] & – & 29.53 [7] & -2.68 [9] & – & 3.94[19] & – \\ \hline
L & Proxima Cen & 4850 [30] & 1.82 [65] & 2.7 [66] & 0.003 [14] & –4.29 [67]& 26.82 [68] & -3.98 [68] & 200 [69]& 89.8[69] & 0.63 [69] \\ \hline
S &  & 4600 [21] & 0.65 [52] & 1.7 [21] & 1 & -4.9$^{+0.15}_{-0.2}$ [53] & 27.2$_{-0.61}^{+0.39}$ [55] & -6.39$^{+0.61}_{-0.49}$ [56] & 5 & 25 [21] & 2 \\ \hline
\end{tabular}%
\vspace{0.12cm}
\caption{Activity indicators for the ALMA-EMISSA sample stars. These include age, B-V, rotation speed, bolometric luminosity, Log $R^\prime_\mathrm{HK}$, X-ray luminosity ($L_\mathrm{X}$), $R_\mathrm{X}$ ($L_\mathrm{X}/L_\mathrm{bol}$), magnetic field strength, rotation period, and Rossby number (\Ro). The quantities marked by [$*$] are estimated using inter-property correlation functions given by \cite{stepien94_Defn_Rx+Ro_Vs_activity_n_manyCorCurves}.\protect\\
{\bf References}: 1: \cite{Grether06_Ref1}, 2: \cite{Wright04_Ref2}, 3: \cite{Soubiran16_Ref3}, 4: \cite{Prieto99_Ref4}, 5: \cite{Mermilliod92_Ref5}, 6: \cite{Messina17_Ref6}, 7: \cite{Malo14_Ref7}, 8: \cite{Vidotto14_B_Vs_age_n_rot}, 9: \cite{Riaz06_Ref9}, 10: \cite{Wright11_Ref10}, 11: \cite{Mamajek08_Ref11}, 12: \cite{Satssun19_Ref12}, 13: \cite{Schmitt04_Ref13}, 14: \cite{Gaia18_DR2_catalog}, 15: \cite{Ramirez12_Ref15}, 16: \cite{Raghavan10_Ref16}, 17: \cite{Takeda07_Ref17}, 18: \cite{Suchkov03_Ref18}, 19: \cite{Gunther20_Ref19}, 20: \cite{2009MNRAS.394.1338B}, 21: \cite{2014MNRAS.444.3517M}, 22: \cite{2015ApJ...804..146D}, 23: \cite{2012AJ....143..135V}, 24: \cite{1975A&AS...22..239N}, 25: \cite{2013ApJ...764...78R}, 26: \cite{1980AcA....30..453G}, 27: \cite{2012A&A...542A.116A}, 28: \cite{2009A&A...501..941H}, 29: \cite{2001A&A...373..159C}, 30: \cite{2002A&A...392L...9T}, 31: \cite{2007A&A...470..295B}, 32: \cite{2010ApJ...722..343D}, 33: \cite{2012A&A...537A.147H}, 34: \cite{2000A&A...355L..27H}, 35: \cite{Ujjwal20}, 36: \cite{Mamajek12}, 37: \cite{DiFolco04}, 38: \cite{2007A&A...474..653V}, 39: \cite{Mora01}, 40: \cite{Luck17}, 41: \cite{Montesinos16}, 42: \cite{Gray03}, 43: \cite{2017ApJ...840...70B}, 44: \cite{Liebert05}, 45: \cite{Royer02}, 46: \cite{Mamajek12}, 47: \cite{2009ApJ...705...89L}, 48: \cite{2011ApJ...734...67M}, 49: \cite{2014AJ....148...54H}, 50: \cite{2018A&A...616A.108B}, 51: \cite{2017MNRAS.469.3042N}, 52: \cite{1992PASP..104.1035G}, 53: \cite{1996AJ....111..439H}, 54: \cite{1997MNRAS.284..202A}, 55: \cite{2015A&A...577L...3T}, 56: \cite{2010PNAS..107.7158T}, 57: \cite{2007AN....328.1037F}, 58: \cite{2007A&A...475..677S}, 59: \cite{2002A&A...384..180F}, 60: \cite{2012A&A...537A.120Z}, 61: \cite{2017ApJ...836..139F}, 62: \cite{2016ApJ...826..171G}, 63: \cite{2018A&A...614A.125F}; 64: \cite{2017A&A...598L...7K}, 65: \cite{2014AJ....147...21J}, 66: \cite{2006A&A...460..695T}, 67: \cite{Sreejith20_RHK_indices}, 68: \cite{Magaudda20_XrayActivity_dMs}, 69 \cite{2021MNRAS.500.1844K}
}
\label{tab:activityprops}
\end{sidewaystable}

%
\begin{table*}[ht!]
\begin{center}
\begin{tabular}{|c|c|c|c|c|c|}
\hline
\multirow{2}{*}{\textbf{ID}} & \multicolumn{1}{c|}{\multirow{2}{*}{\textbf{SIMBAD ID}}} & \textbf{ALMA Obs. frequency} & \textbf{Other bands} & \multirow{2}{*}{\textbf{Refs}} & \multirow{2}{*}{\textbf{$\beta$}} \\ \cline{3-4}
 & \multicolumn{1}{c|}{} & (GHz) & Array: (GHz) &  &  \\ \hline
A & Alf CMa & 94, 106, 144,156 196, 210 & VLA: 33, 45;  GBT: 90 & 11 & 1.8$^*$ \\ \hline
B & Alf PsA & 233, 344 & — & 9 & 2.07$^*$ \\ \hline
C & Gam Vir A & 233, 344 & — & 1 & 1.87$^*$ \\ \hline
D & Gam Vir B & 233, 344 & — & 1 & 1.87$^*$ \\ \hline
E & Eta Crv & 341 & — & 10 & — \\ \hline
F & Gam Lep & 232, 345 & — & 1 & 1.87$^*$ \\ \hline
G & Alf Cen A & 98, 145, 233, 344, 405, 679 & ATCA: 17 & 3, 4 & 1.76$^3$ \\ \hline
H & 61 Vir & 341 & — & 2 & — \\ \hline
I & Alf Cen B & 98, 145, 233, 344, 405, 679 & ATCA: 17 & 3, 4 & 1.71$^3$ \\ \hline
J & Eps Eri & 228 & VLA: 6, 10, 15, 33;  ATCA: 44 & 5, 6 7, 8 & — \\ \hline
K & GJ 2006 A & 341 & — & — & — \\ \hline
L & Proxima Cen & 233 & -- & 12 & — \\ \hline
S & \multicolumn{1}{c|}{Sun} &92-108, (190-206), 229-249, (339-355)
 & VLA: 10; NoRH: 17, 34 & 13, 14 & -- \\ \hline
\end{tabular}%
\caption{Details of the available mm -- cm band data for the sample stars. Columns 3-4: The frequencies at which significant stellar detections were made so far in mm - cm band. Column 5: References to publications that reported the data. Column 6: The ALMA band spectral index either derived from JVO images or reported in the literature. 
The spectral index $\beta$ is given for stars for which reliable flux values are available for more than one frequency in 10 -- 1000\,GHz. $\beta$ is not provided for the Sun because a single value cannot be defined for the solar mm -- cm spectrum. 
\textbf{References:} 1: \cite{White20_MESAS}, 2: \cite{2017MNRAS.469.3518M}, 3: \cite{2019arXiv190403043L}, 4: \cite{2018MNRAS.481..217T}, 5: \cite{2019ApJ...871..172R}, 6: \cite{Bastian18_EpsEri}, 7: \cite{2015ApJ...809...47M} , 8: \cite{2017MNRAS.469.3200B}, 9: \cite{2016ApJ...818...45S}, 10: \cite{2017MNRAS.465.2595M}, 11: \cite{2019ApJ...875...55W}, 12: \cite{MacGregor18_proxima_flares}, 13: \cite{2017SoPh..292...88W}, 14: \cite{2004NewAR..48.1319W}.  
}
\label{tab:Radio_band_info}
\end{center}
\end{table*}


{In this section, we will first describe the compilation of ALMA detected stellar sample and the subsequent search for their flux densities reported at different frequencies. We will then present the derivation of their expected flux densities across the electromagnetic spectrum using a photospheric emission model. The model flux densities are compared with the observed values by defining a suitable parameter.}

\subsection{Compilation of the initial stellar sample}
\subsubsection{Archival data in the mm--cm band ($\approx$ 10 -- 1000\,GHz)}
\label{sec:archive_radiomm}

As a starting point for the ALMA main-sequence star sample, the ALMA Science Archive\footnote{\href{https://almascience.eso.org/asax/}{https://almascience.eso.org/asax/}} was searched for significant  detections of main-sequence stars.  
A detection was deemed significant if  
(i)~the flux density of the target star was beyond at least three times the root mean square (RMS) flux of the background sky in the image, and 
(ii)~there was no nearby or background source within the synthesised beam or close enough to contaminate the observed stellar flux. 
The primary criterion for a significant detection placed a strong constraint on the distance to the stars, making it a collection of nearby main-sequence candidates. 
We refer to Sect.~\ref{sec:disc_samplesize} for a discussion of the current limitations of the sample size and future possibilities to expand it.
 
ALMA observations are generally classified under various science categories\footnote{\href{https://almascience.eso.org/alma-science}{https://almascience.eso.org/alma-science}} according to their intended primary scientific goal. Here, data sets under the following categories were searched for detectable main-sequence stars: `Debris disks', `Exoplanets', `Disks around low-mass stars', `Disks around high-mass stars', and `Main-sequence stars'. 
In total, publicly available data sets for 104 main sequence stars were identified. From these, 8 stars with unknown spectral type or luminosity class and 4 stars in spectroscopic binaries were removed. This led to a reduced sample of 92~stars. The various observed and derived physical information of these stars was tabulated using an automated search routine written in Python, which queried the Vizier and SIMBAD databases. 

After this preliminary sample was ready, the first-look images of these stars from the various observation campaigns were downloaded from the ALMA Virtual Observatory (VO) through the Japanese Virtual Observatory\footnote{\href{https://jvo.nao.ac.jp/portal/alma.do}{https://jvo.nao.ac.jp/portal/alma.do}} (JVO) website. The signal-to-noise ratio (S/N) of each image was noted, and those with a significant stellar detection were downloaded. Out of the 92~candidate stars, data for many were found to have contamination from nearby sources, which was visible from the shape of the stars, which was significantly different from that of the synthesised beam. In many other cases, there was either no detection of the star itself or data were poor. The final sample of stars with a significant detection in at least one ALMA band consists of 12 sources. 
The sample is complete for ALMA data that are publicly made available through JVO until 31 December 2020 since the beginning of observations.
Figure~\ref{fig:Sirius_imgs} shows a sample collage of JVO-supplied ALMA images of $\alpha$\,CMa (Sirius A). 

The flux densities of different ALMA bands for each star were determined by fitting a two-dimensional Gaussian function to each image downloaded from the JVO. 
A detailed search in the VLA data archive \footnote{\href{https://archive.nrao.edu/archive/advquery.jsp}{https://archive.nrao.edu/archive/advquery.jsp}} was also carried out to identify archival data sets for the stars in the sample. All the identified data sets were imaged using an automated pipeline that we developed based on routines in Common Astronomy Software Applications (CASA; \citealp{casa}). Every sample star except for GJ\,2006\,A was found to have publications using at least a few of the mm -- cm band detections we found from the archival search. 

ALMA flux values from publications were preferred over values derived from JVO images because the latter are based on a standard pipeline, while flux values in publications are typically the result of more refined data-processing optimised for an individual star. However, the published fluxes from stellar observations with the VLA match our pipeline values quite well. A few stars had published ATCA observations as well, which are incorporated in our database.

\subsubsection{Archival data at higher frequencies}
\label{sec:archive_highfreq}

The 10 -- 1000\,GHz band information (see Sect.~\ref{sec:archive_radiomm}) were complemented with data at higher frequencies ranging from the infrared (IR) to the EUV. For this purpose, the  
Strasbourg Data Centre\footnote{\href{http://vizier.unistra.fr/vizier/sed/}{http://vizier.unistra.fr/vizier/sed/}} (CDS)
was searched for each star in the sample. 
When provided the RA and DEC of a target source and a search radius, the CDS server looks for sources detected within the radius across the different astronomical catalogues and reports their fluxes as a function of frequency. We used this tool to obtain the SEDs of the stars of interest. A caveat in this procedure is the choice of an optimal search radius. CDS by default sets a radius of 5$^{\prime\prime}$. 
Three criteria that determine this choice are the source density in the region of interest, the presence of known associated sources of emission, such as binary companions and disks near the stars of interest, and to have sufficient data points in the SED given the typical pointing offsets of various telescopes.

The default search radius of $5^{\prime\prime}$ is well justified considering all three aforementioned criteria for all pre-selected stars except for $\gamma$ Vir A\&B, which are separated by just \asecdot{2}{6}  \citep{1999A&A...341..121S} or less, depending on the time of observation during the past decades. %
Most of the data points provided by CDS were obtained with instruments with an angular resolution $\gtrsim 2^{\prime\prime}$. Especially in the far-IR bands ($\sim 10^3 - 10^4$\,GHz), the data were obtained with instruments at a resolution  $>5^{\prime\prime}$.
For instance, Spitzer (MIPS\footnote{\href{https://irsa.ipac.caltech.edu/data/SPITZER/docs/mips/mipsinstrumenthandbook/3/}{https://irsa.ipac.caltech.edu/data/SPITZER/docs/mips/}}
) offers resolution 6$^{\prime\prime}$, 20$^{\prime\prime}$, and 40$^{\prime\prime}$ at wavelengths of 24\,$\mu$m, 70\,$\mu$m, and 160\,$\mu$m (corresponding to frequencies of $\approx 12.5$\,THz, 4.3\,THz, and 1.8\,THz), respectively.
The surveys at frequencies between  $\approx 10^4$ -- $4\times 10^5$\,GHz, namely 2MASS\footnote{\href{https://irsa.ipac.caltech.edu/data/2MASS/docs/supplementary/seeing/seesum.html}{https://irsa.ipac.caltech.edu/data/2MASS/docs}}
and Spitzer \citep[IRAC;][]{2004ApJS..154...10F}, offer $\approx$ 2 -- $\asecdot{2}{5}$ pixel scales.  
As a result, flux contamination is inevitable in the CDS SED data for $\gamma$\,Vir\,A and B at frequencies below $10^5$\,GHz.
However, for frequencies above $5\times 10^5$ (optical to UV), the angular resolution of the available data, which mostly stems from catalogues like Gaia \citep{Gaia18_DR2_catalog}, {is about a few} to sub-arcseconds.
Consequently, we used a CDS search radius of size $\asecdot{1}{5}$ for both $\gamma$\,Vir\,A \& B and then corrected the fluxes in the range from $\nu = 10^3$\,GHz to $\nu =  5\times 10^5$\,GHz. 
For this frequency range, it is assumed that both stars contribute to the measured fluxes. Separate fluxes for $\gamma$\,Vir\,A and B were then derived according to the ratio of their \teff\ values as obtained from \citet{Soubiran16_Ref3} (see Table~\ref{tab:prop}). Details of the SED data processed by this approach are discussed in detail in the following section.

\subsection{PHOENIX model atmospheres}
\label{sec:sedmodelling}

\begin{figure*}[hpt]
\centering
\includegraphics[scale=0.2,width=0.94\textwidth,height=0.32\textheight]{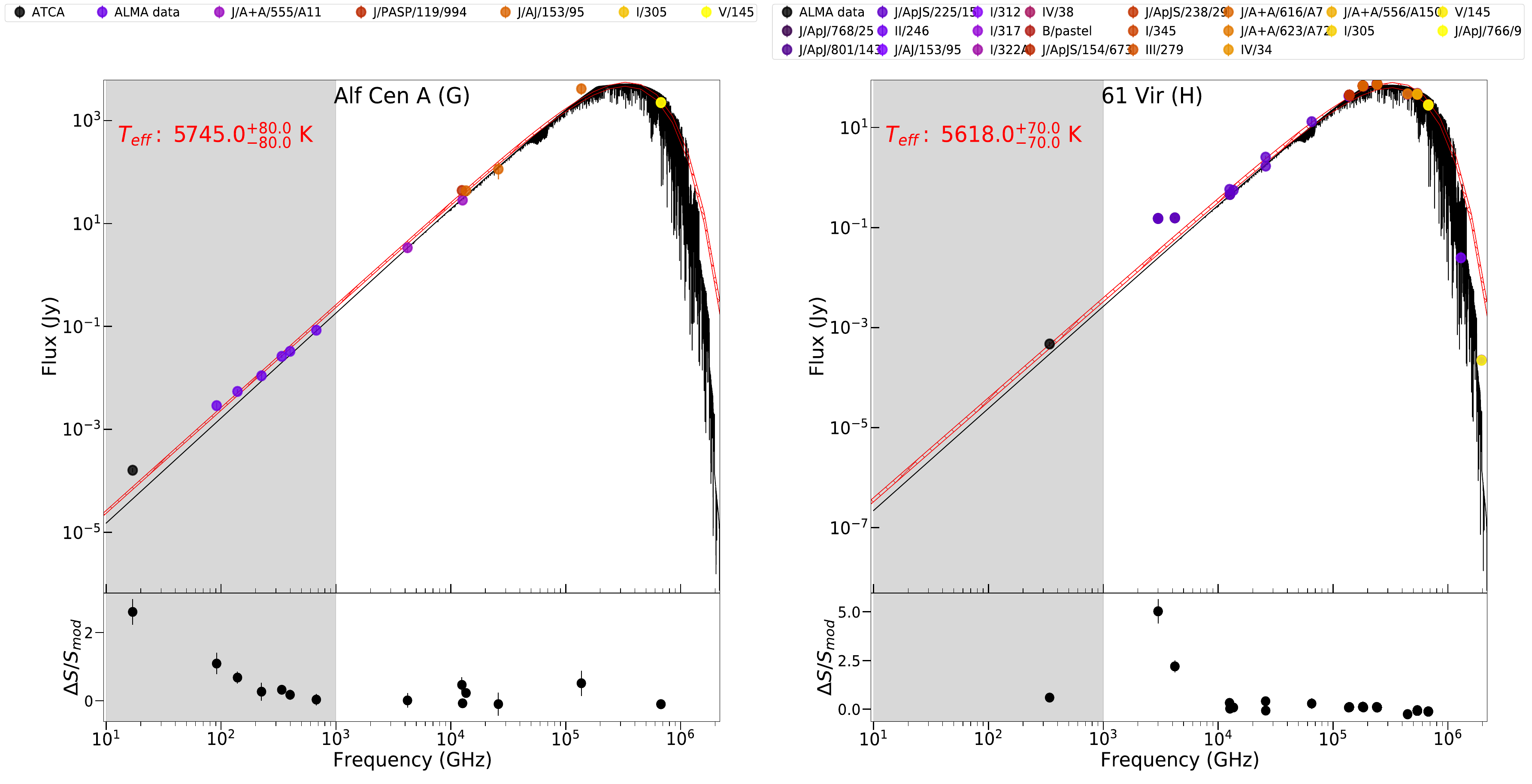}
\includegraphics[scale=0.2,width=0.94\textwidth,height=0.32\textheight]{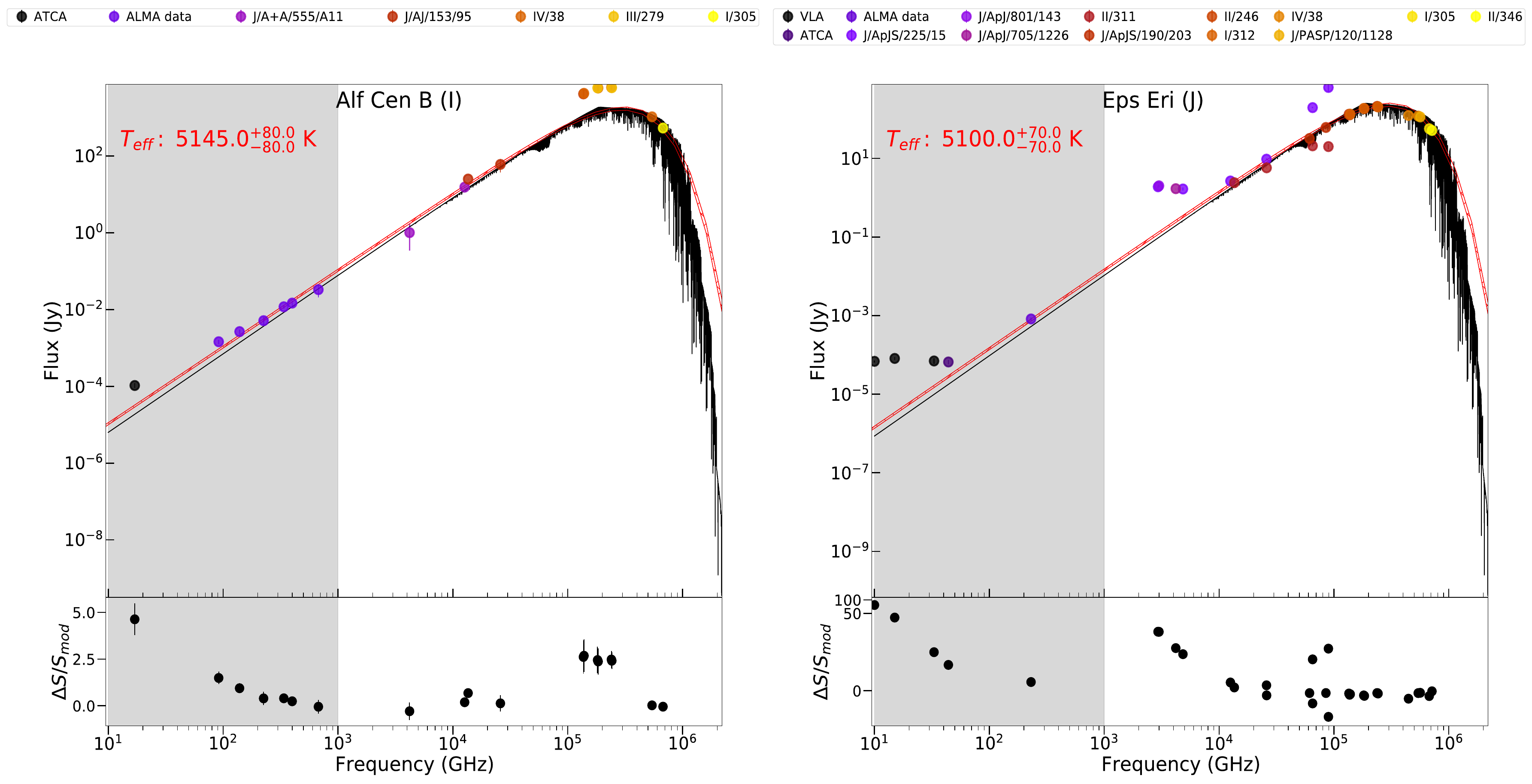}
\includegraphics[scale=0.2,width=0.94\textwidth,height=0.32\textheight]{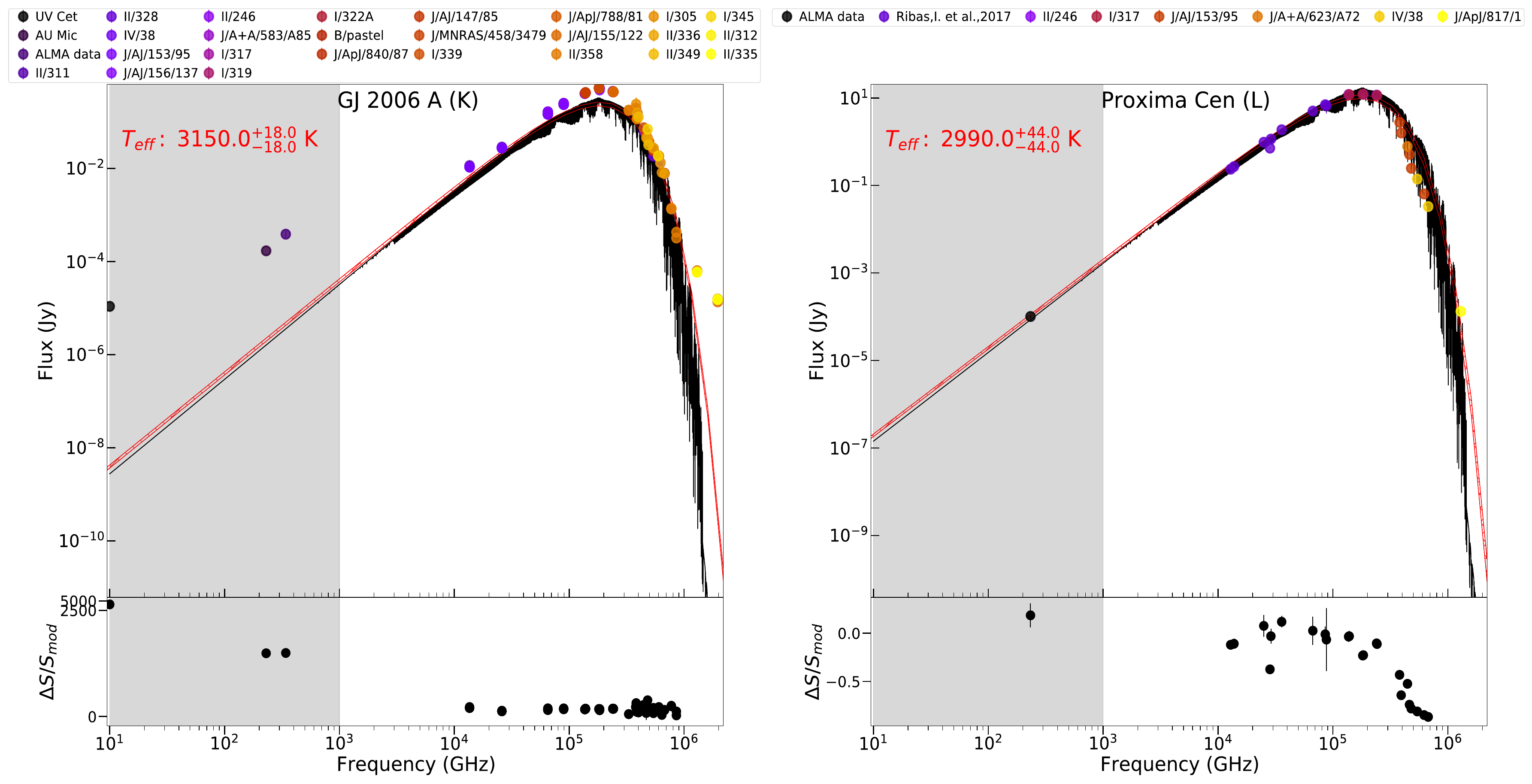}
\vspace{-0.2cm}
\caption{
  SED data for every star shown with legends providing VizieR catalogue references. The red meshed region shows a blackbody model with \teff\ values from \protect \cite{Soubiran16_Ref3}, specified in each panel. The black curve shows the PHOENIX SED model.
}
\label{fig:SED_fits}
\end{figure*}

A one-dimensional local thermal equilibrium (LTE) model atmosphere and the corresponding synthetic spectrum were calculated for each star in the sample using the PHOENIX code \citep{1999JCoAM.109...41H} with the model grid taken from \cite{2013A&A...553A...6H} as the starting points with current spectral line and thermodynamic data. 
For the stellar parameters, rounded literature values were used (see Sect.~\ref{sec:results}).
The model atmospheres feature a photosphere in radiative + convective equilibrium for the stellar parameters using the current production version of PHOENIX (version 18.05).
Consequently, none of the models includes a chromospheric temperature rise. 
The synthetic spectrum covers a wavelength range from 0.01\,nm to 3.0\,cm, corresponding to frequencies from $3.0\,10^{10}$ to 10\,GHz. 
Towards the lower frequencies, the synthetic SEDs ($S_\mathrm{mod}\,(\nu)$) follow the Rayleigh-Jeans tail.  
In this study, the PHOENIX spectra are compared to the observed SEDs, which will then expose additional contribution due to a chromosphere and/or possible circumstellar disks (see Sect.~\ref{sec:results}). 
The resulting SEDs are presented in Fig.~\ref{fig:SED_fits}. 

\label{sec:fluxexcessparameter}
All considered stars were compared to corresponding PHOENIX models that do not include a chromospheric temperature rise. Deviations between the PHOENIX model SED ($S_\mathrm{mod}(\nu)$) and the observed values ($S_\mathrm{obs}(\nu)$) are discussed with the help of the {\em \textup{spectral flux excess}} parameter, which is defined as 
\begin{equation}
    \Delta S(\nu)/S_\mathrm{mod} = \frac{S_\mathrm{obs}(\nu)- S_\mathrm{mod}(\nu)}{S_\mathrm{mod}(\nu)} \ . 
    \label{Eqn:DelS}
\end{equation}
A significant spectral flux excess in the mm -- cm range would potentially indicate a chromospheric temperature rise because the 
PHOENIX model lacks atmospheric layers above photosphere.

\begin{figure*}[tph]
\centering
\includegraphics[scale=0.2,width=0.95\textwidth,height=0.66\textheight]{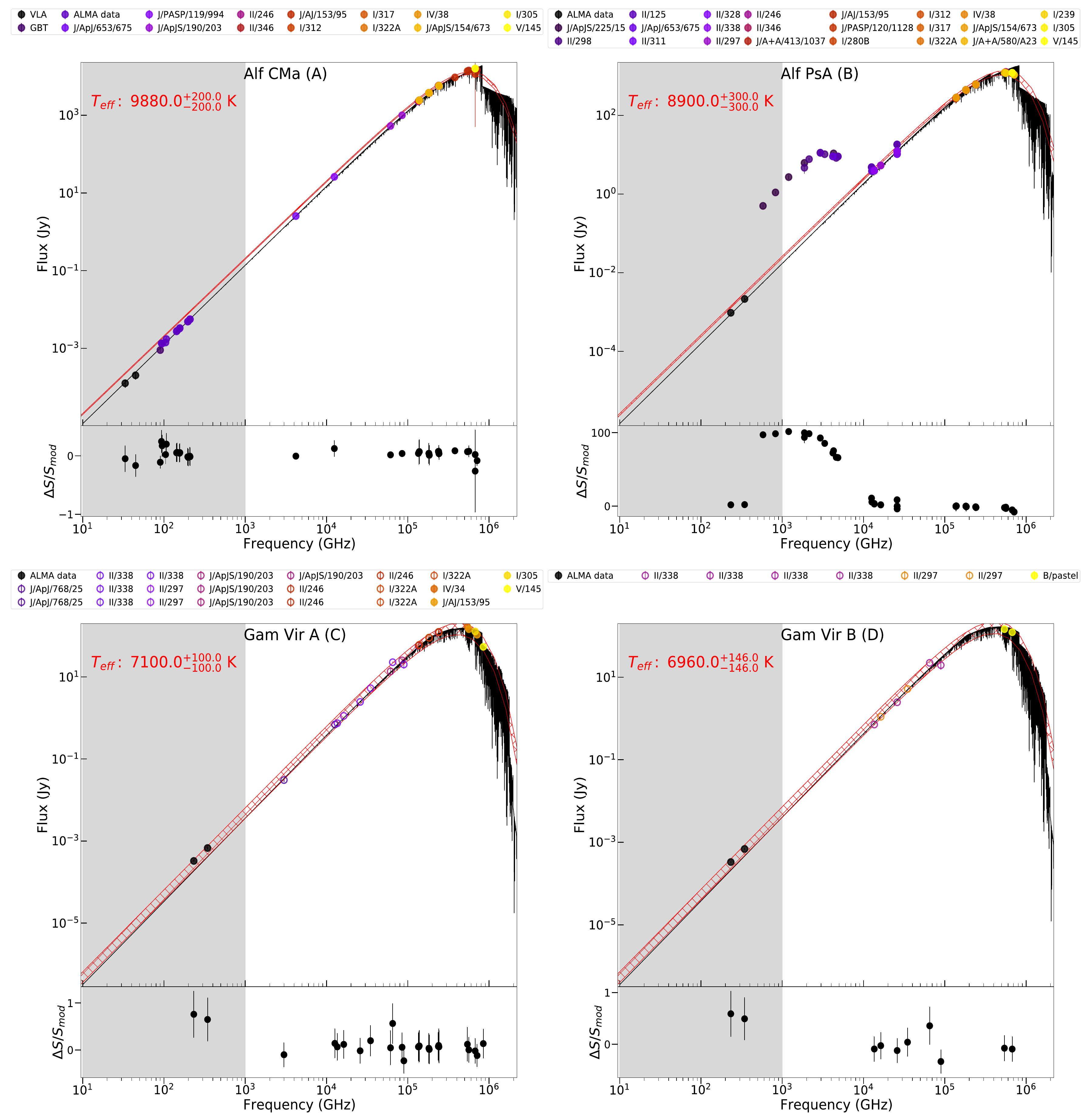}

\includegraphics[scale=0.2,width=0.96\textwidth,height=0.32\textheight]{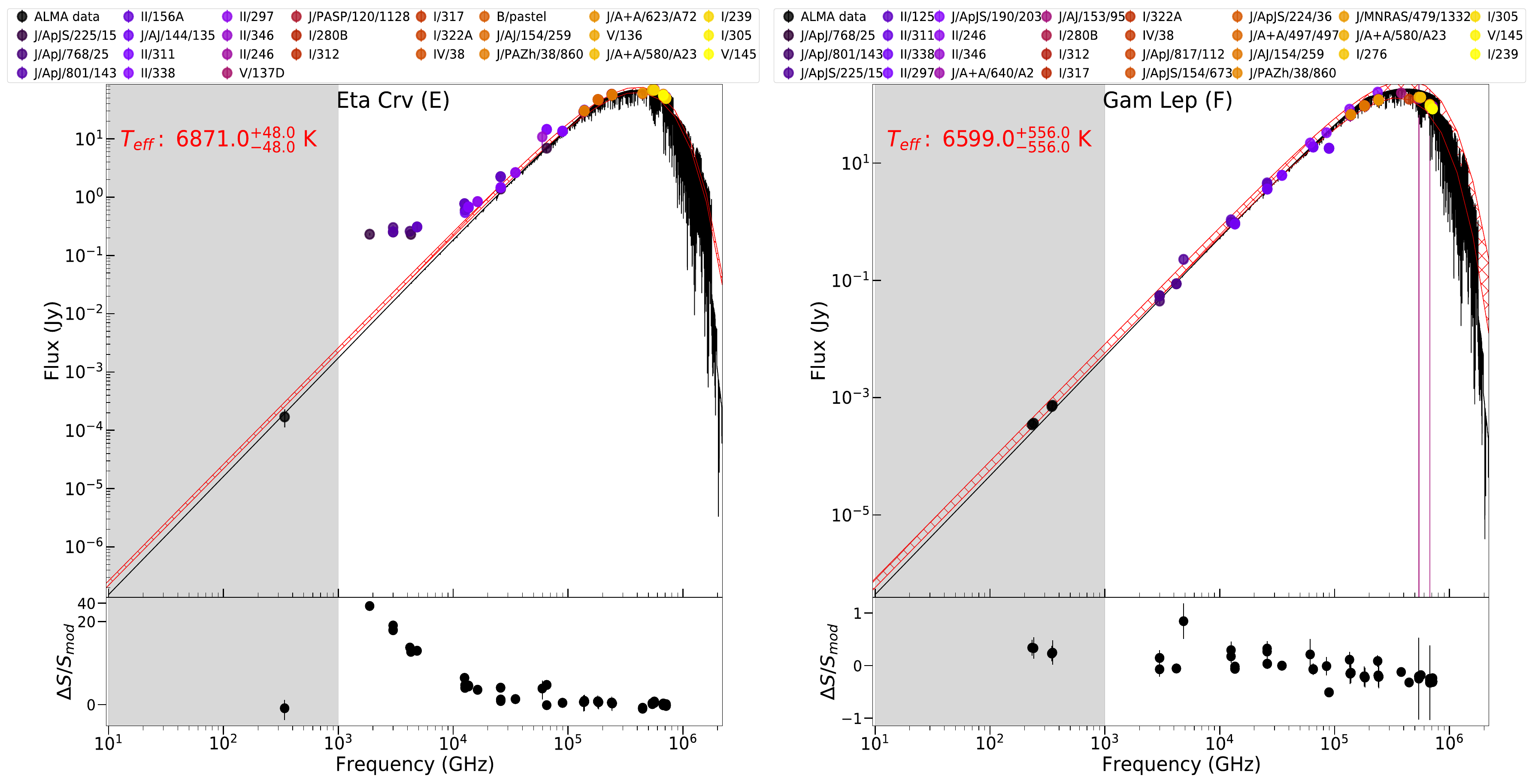}
\end{figure*}

\section{Results}
\label{sec:analysis}
\label{sec:results}
\begin{figure}[th!]
    \centering
    \includegraphics[width=8cm]{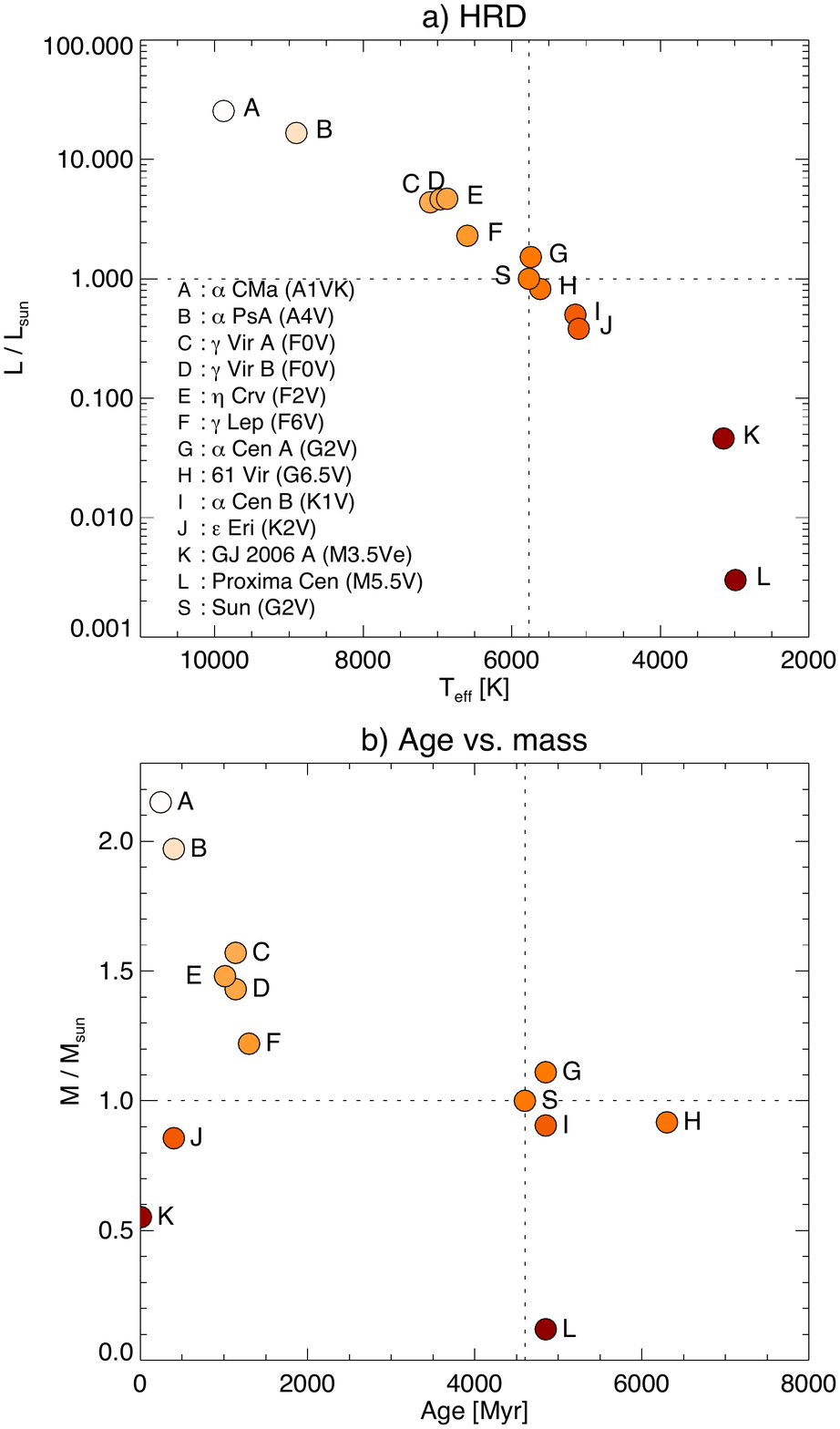}
     \caption{Distribution of the stars in the sample: \textbf{a)} Hertzsprung-Russell diagram of the sample showing \teff\ and luminosity. \textbf{b)}~Age vs. mass. We refer to Tables~\ref{tab:prop}-\ref{tab:activityprops} for the values and references.} 
    \label{fig:hrd}
\end{figure}
The resulting initial ALMA main-sequence star sample is comprised of 12~stars that satisfy our selection criteria plus the Sun as a reference star. 
A future expansion of this list is foreseen (see Sect.~\ref{sec:disc_samplesize}). 
Figure ~\ref{fig:hrd} presents the covered parameter ranges across the main sequence and stellar age versus mass. 
An overview of the physical parameters and activity indicators for each star in the sample and the Sun is given in Table~\ref{tab:prop}-\ref{tab:activityprops}. 
In Table~\ref{tab:Radio_band_info},  the observation frequencies in the mm -- cm range are presented for which reliable data exist for each star.  
For the stars with reliable flux estimates \Snu\ at more than one frequency, a spectral index, $\beta$ was computed, assuming \Snu\,$\propto \nu^{\,\beta}$.
The results of this analysis for each star are discussed here starting with our template star, the Sun. 

\subsection{Sun as a star.}  
The Sun is an intermediate-age ($\approx$ 4.6\,Gyr) main-sequence  G2V star with an effective photospheric blackbody temperature of 5772\,K \citep{2014MNRAS.444.3517M}.  The Sun has an activity cycle of 11\,yr and a (equatorial) rotation period of about 25\,days. The \RHK\ index of the Sun is known to fluctuate by about 60\% during its activity cycle (see Table~\ref{tab:activityprops}).
This level of fluctuation is common for a star at the age of the Sun, as demonstrated by \cite{Pace13_RHK_Drastic_variability}.

\begin{figure}
    \centering
    \includegraphics[width=0.45\textwidth,height=0.35\textheight]{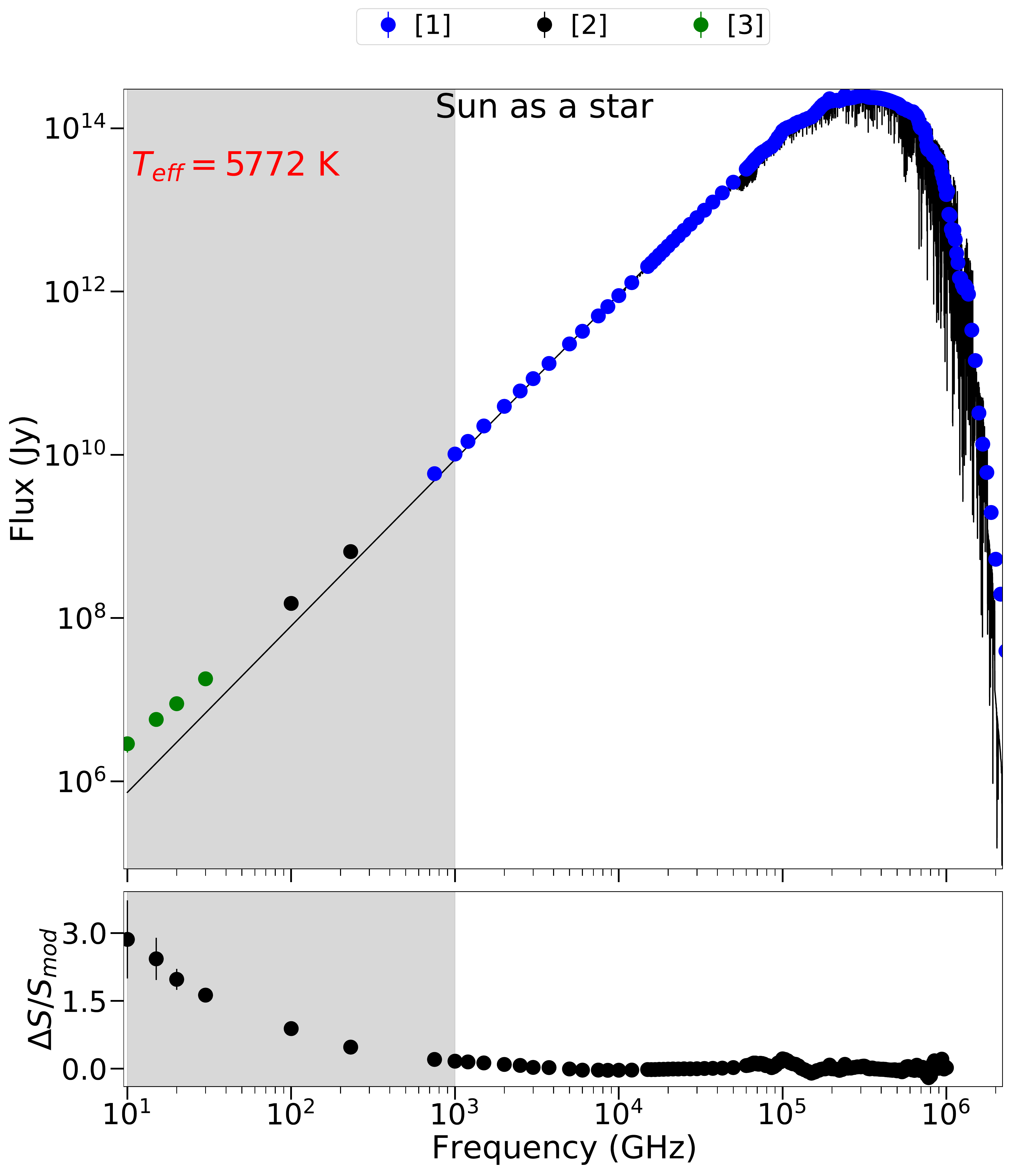}
    \caption{SED for the Sun as a star. A blackbody spectrum assuming $T_{eff} = 5772\,K$
    is shown. References for the solar data: 1: \citet{Thekaekara69}, 2: \citet{2017SoPh..292...88W}, and 3: \citet{2004NewAR..48.1319W}.
    }
    \label{fig:SolarSED_fits}
\end{figure}
Despite the wealth of data across spectral bands, still comparatively little data are available in mm -- cm bands. 
The full-disk averaged SED data in the spectral range from $\approx 10^3$\,GHz to  $2\times 10^6$\,GHz ($\approx 150$\,nm -- 0.5\,mm) for the Sun shown in Fig.~\ref{fig:SolarSED_fits} are taken from \cite{Thekaekara69}. Despite being a relatively old SED data set, it provides data over a very wide spectral range compared to other more recent data sets. 
Using an archival solar irradiance data set compiled over several decades, \cite{2019E&SS....6.2525C} found that the relative variance in the spectral irradiance is lower than 0.1\% in the spectral range from $400$ -- 2000\,nm and lower than 0.7\% in the 200 -- 400\,nm range.
Consequently, the spectrum provided by \cite{Thekaekara69} still serves as a valid reference for the disk-averaged solar photospheric spectrum. 

\cite{2017SoPh..292...88W} provided the fluxes based on ALMA total power solar maps in bands 3 and 6, that is, at 92-108 and 229-249\,GHz, respectively.
Fluxes at $\nu < 30$\,GHz were estimated based on the disk-averaged fluxes computed by \cite{2004NewAR..48.1319W} using archival data from the Nobeyama Radio Heliograph\footnote{\href{http://solar.nro.nao.ac.jp/norh/}{http://solar.nro.nao.ac.jp/norh/}}. 
The author provided a Sun-as-a-star spectrum for the maximum and minimum phases of the solar activity cycle based on data collected during December 2001 and May 1996, respectively.
The data presented in Fig.~\ref{fig:SolarSED_fits} are mean values computed from the two phases, and the bars mark the flux range arising from variations, within $\sim$ 1 -- 20\%, due to solar activity. 
Updated fluxes for the Sun at millimeter wavelengths will be addressed in a forthcoming paper in this series.
The lower subplot of Fig.~\ref{fig:SolarSED_fits} shows the spectral flux excess, \DSmod, for the Sun as a star. 
The PHOENIX model for the Sun is overlaid.
\DSmod\ steadily increases with decreasing frequency in the mm -- cm band, while it remains relatively close to 0 at high frequencies between 10$^3$\,GHz and 10$^6$\,GHz, where the emission primarily originates from the photosphere. 
The rise in spectral flux excess with decreasing frequency in the mm -- cm range indicates that these frequencies are sensitive to atmospheric layers of varying heights above the photosphere, extending across the chromosphere to corona.
The solar observations in the sub-mm -- far-IR range ($\approx$ 300 -- 10$^4$\,GHz)  show that the solar spectrum has a \Tb\ minimum of $\approx$ 4100\,K at $\sim \nu \approx$ 2000\,GHz ($\sim 150~\mu$m)  \citep[e.g.][]{1981ApJS...45..635V,1997ApJ...484..960G}.
The early millimeter observations were mostly solar fluxes measured at the disk centre or scans across the disk centre. In some cases, these measurements were obtained with airborne instruments. However, these measurements had significant uncertainties compared to modern interferometric observations. The average flux values derived from measurements for which only parts of the Sun and not the full disk were observed would have been more susceptible to contributions from structures on the solar surface during the observation time.  
In this work, we compiled only data from more sensitive mm -- cm band  interferometric observations and full-disk scans/total power maps \citep{2017SoPh..292...88W}.

\subsection{Stellar sample}
\label{sec:stellarsample}
{Short descriptions of the individual stars are provided below. The stars are presented in the decreasing order of their \teff.}

\begin{figure*}
    \centering
    \includegraphics[width=\textwidth]{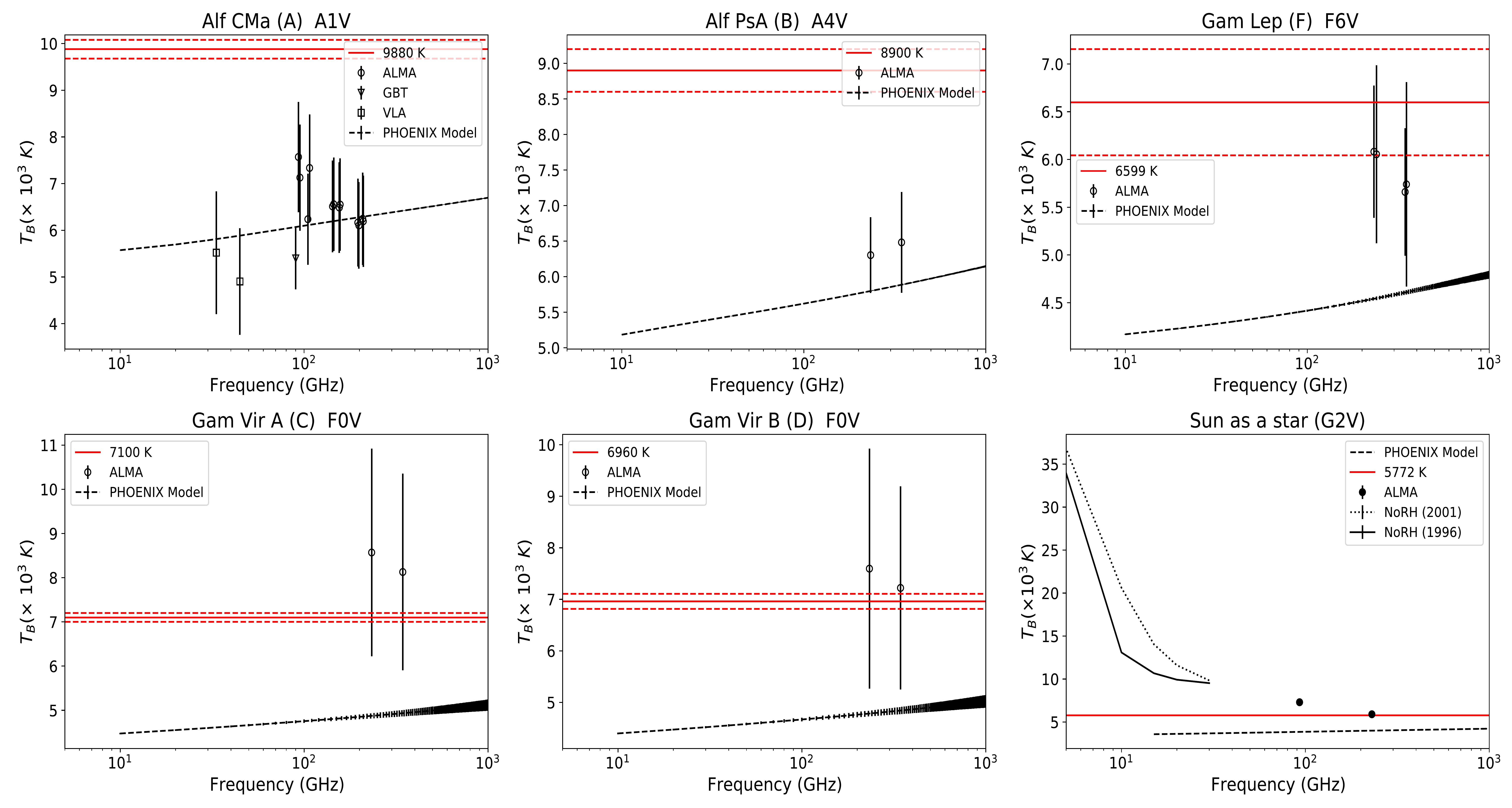}
    \includegraphics[width=\textwidth]{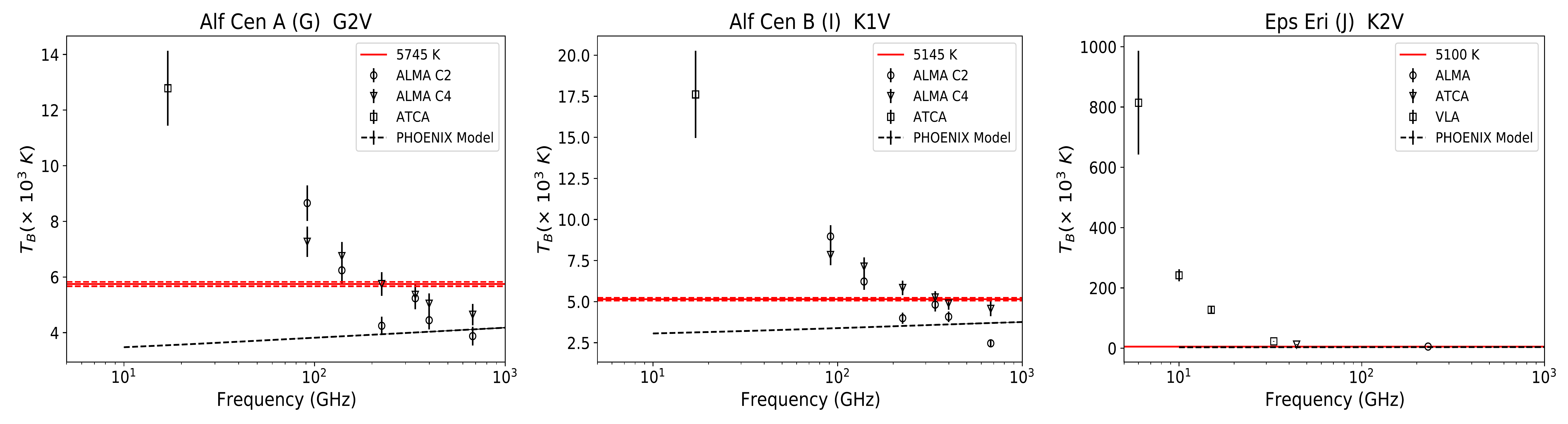}
    \caption{{\it Left to right}: T$_B(\nu)$ for stars including Sun as a star in decreasing order of spectral type. The \Tbp\ for the Sun during solar maximum (December 2001) and minimum (May 1996) are shown separately. PHOENIX model T$_B(\nu)$ is shown along with the \teff\ values for each star. The error interval in \teff is marked by a dashed red line.
    }
    \label{fig:tbprofile}
\end{figure*}

\paragraph{A --- $\alpha$\,CMa (Sirius A, HD\,48915A)}  
is a main-sequence star of spectral type A0 and part of a binary system. The companion is a white dwarf (spectral type DA). Sirius is observed spatially separated from its companion at $\approx$ 6$^{\prime\prime}$ -- 8$^{\prime\prime}$ \citep{2017ApJ...840...70B}, and with a distance of 2.6\,pc, it is one of the closest stars to the Sun. It has an age of 242\,Myr and a mass of 2.15\,\Ms.
Available ALMA observations cover bands 3, 4, and 5 (see Table~\ref{tab:Radio_band_info}). 

The SED at mm wavelengths (see Fig.~\ref{fig:SED_fits} top right panel for $\nu< 300$\,GHz) is well reproduced with the PHOENIX model, which confirms the results by \citet{2018ApJ...859..102W}. Like the flux values, the corresponding \Tb values (see Fig.~\ref{fig:tbprofile}) can also be explained with a temperature stratification that features a monotonic decrease with height, as already demonstrated by \citet{2018ApJ...859..102W}. As expected for an A0 star, there is no indication of a chromospheric temperature rise. The \Tb data indicate some potential deviation from a monotonic temperature stratification between $\sim40$\,GHz and $\sim80$\,GHz in the form of an unexpected dip, but more flux data are needed before firm conclusions can be drawn.

\paragraph{B --- $\alpha$~PsA (Fomalhaut, HD\,216956)}  
is an A4V star situated at a distance of 7.7\,pc. It has a mass of 1.97 M$_\odot$ and an age of 0.44\,Gyr. $\alpha$\,PsA forms a trinary system with HD\,216803 and the distant LP 876-10, both more than a few degrees away \citep{Mamajek13_fomalhaut_3rd_companionDiscovery}. ALMA has resolved the thick elliptical debris ring around $\alpha$\,PsA. The inner edge of the closest ring  to the star is located at about 3$^{\prime\prime}$ away from the star \citep{2017ApJ...842....8M}. The ring has a thickness of $\sim$\,2$^{\prime\prime}$. 
The available ALMA observations are in bands 6 and~7. 
These two data points (see frequencies lower than 400\,GHz in Fig. \ref{fig:SED_fits} and Fig.~\ref{fig:tbprofile}) match the fluxes (and \Tb) predicted by the PHOENIX model. 
An increased flux is found between 400\,GHz and $<10$\,THz. These data come from the HERSCHEL SPIRE spectrograph, which has a resolution of $\approx$ 20$^{\prime\prime}$. So the observed flux excess could be attributed to contamination from the debris disk. 
However, 70 $\mu$m Herschel PACS image of the star with an angular resolution of $\asecdot{5}{6}$ could barely distinguish the emission from outer regions of the debris ring, despite the contamination to starlight from the inner regions \citep{2012A&A...540A.125A}. 
As expected for an A-star, the SED for $\alpha$~PsA does indicate the lack of a chromosphere and thus no temperature increase in the outer layers. 

\paragraph{C+D --- $\gamma$~Vir A \& B (HD\,110379, HD\,110380)}  are two very similar F0V stars forming a close binary system at a distance of 11.62\,pc. Almost all their important physical properties match within the respective measurement errors. Their binary orbit is inclined at 148$^\circ$, has a semi-major axis extent of $\asecdot{3}{68,}$ and a period of $\approx$\,169~days \citep{1999A&A...341..121S}. During the ALMA observations in bands 6 and 7, the relative separation of the two stars was $\approx$ $\asecdot{2}{6}$, so that they were resolved \citep{White20_MESAS}.  
However, as mentioned in Sect.~\ref{sec:archive_highfreq}, the binary separation of $\gamma$~Vir A\&B is either smaller than or comparable to the angular resolution scale of various instruments or surveys providing data to CDS, so that only the combined flux of both stars is measured. The individual fluxes for A and B, which were here derived according to their \teff\ ratio in the IR/optical range, agrees well with the PHOENIX models (see Fig.~\ref{fig:SED_fits}).

\begin{figure}
    \centering
    \includegraphics[width=0.5\textwidth]{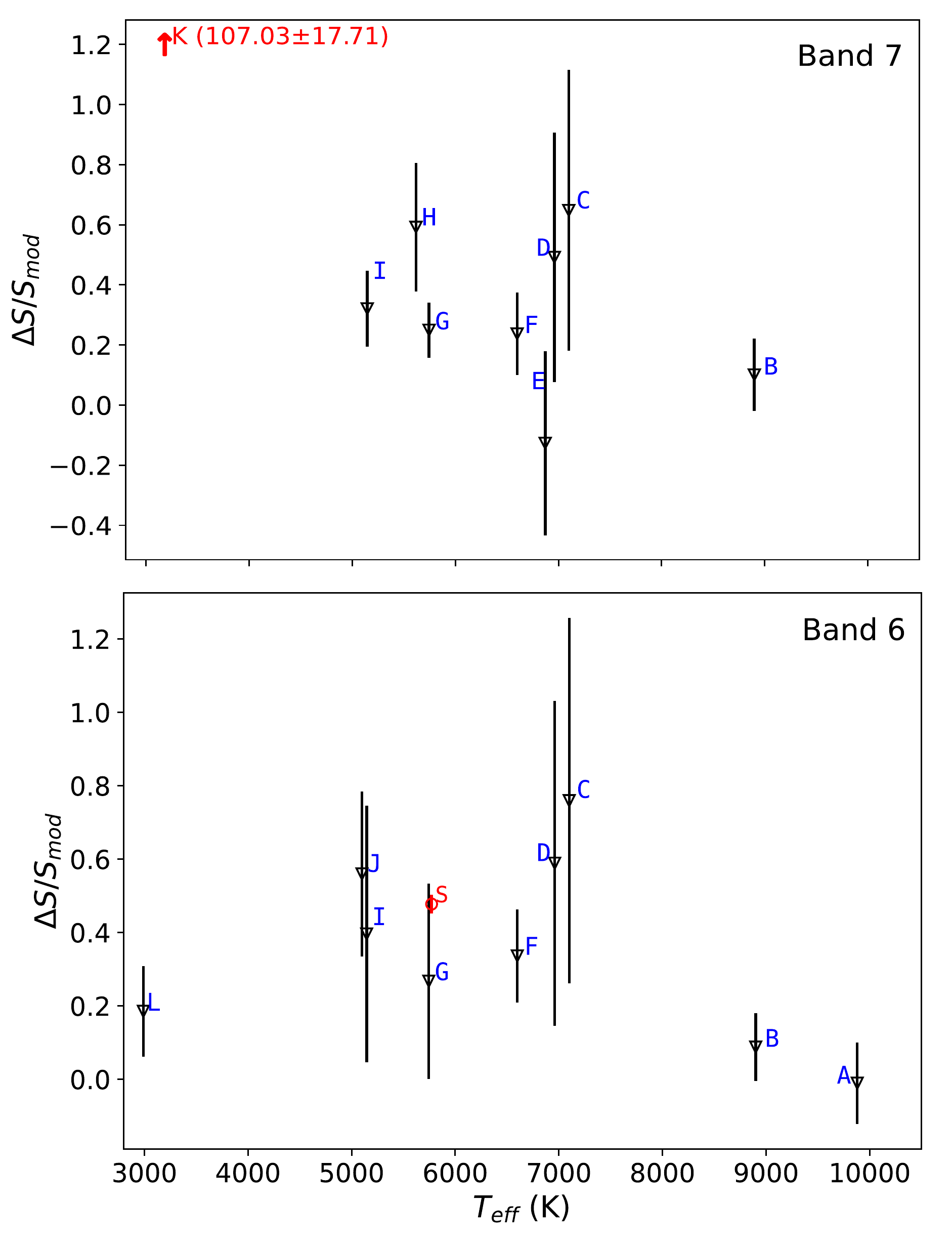}
    \caption{\DSmod\ as a function of \teff\ for band 7 and band 6. The IDs of stars are provided alongside the data points. For GJ 2006 A (ID: K), the value of the spectral flux excess is provided in brackets (top panel).}
    \label{fig:DS_by_S_spectrum}
\end{figure}

Because ALMA separates the two stars well, individual fluxes and corresponding brightness temperatures were derived for the two observed bands. 
For both stars, the ALMA fluxes significantly exceed the fluxes predicted by the PHOENIX model. 
The corresponding brightness temperatures exceed 8000\,K for $\gamma$~Vir~A and 7000\,K for $\gamma$~Vir~B, which is in line with chromospheric temperatures expected for F-type stars (see Fig.~\ref{fig:tbprofile}).
However, it is important to emphasise that uncertainties for the stellar radii, effective temperatures, and to a lesser extent, for their distance, result in a notable flux uncertainty, we show in Fig.~\ref{fig:tbprofile} and Fig.~\ref{fig:DS_by_S_spectrum}. 
In this case, further observations at higher and lower frequencies are also required to determine the average temperature stratification and the properties of a temperature minimum therein.    
\paragraph{E --- $\eta$~Corvi (HD\,109085)}  
is an F2V type star at a distance of 17.96\,pc with a mass of 1.48 M$_\odot$ and an age of 1.01\,Gyr. The star possesses a disk with a semi-major axis width of $\approx$ 17$^{\prime\prime}$, which is well resolved by ALMA in band 7. A detailed modelling of the observed stellar SED assuming multi-thermal emission contributions from the photosphere and a variety of possible hypothetical cold debris disks and rings was carried out by \cite{2014ApJ...784..148D}.
The \DSmod\ plot for $\eta$~Corvi shown in Fig.~\ref{fig:SED_fits} indicates excess flux between 1 --    $10$\,THz. These data come from HERSCHEL SPIRE observations with an angular resolution of about $20^{\prime\prime}$. This should hence be contaminated by the disk emission.
In addition to the outer disk, \cite{2014ApJ...784..148D} also argued for the existence of an inner disk within $\asecdot{0}{6}$ to explain the excess flux observed in the SED at 70\,$\mu$m ($\nu \approx$ 4300\,GHz). However, the contribution from this proposed inner ring to the observed flux is expected to fall below the stellar contribution in the ALMA bands.
The ALMA observations resolve the extended outer disk structure from the star quite well.
The flux measured with ALMA at 341\,GHz, on the other hand, is well reproduced with the PHOENIX model, as is shown by \DSmod\,($\nu=341\,\mathrm{GHz}$)\,$\approx\,0$ in Fig.~\ref{fig:SED_fits}. 
While this finding implies that no chromospheric temperature rise is detected at this frequency, no firm conclusions regarding a potentially existing chromosphere or a lack thereof can be drawn without additional measurements at lower frequencies.

\paragraph{F --- $\gamma$~Lep (HD\,38393)} 
is an F6V type star at a distance of 8.97\,pc with a mass of 1.22\,M$_\odot$ and an age of 1.3\,Gyr. It has a companion at a distance of about 100$^{\prime\prime}$   \citep{White20_MESAS}. ALMA observations are available in bands 6 and~7 and provide fluxes that are higher than those predicted by the PHOENIX model. The \DSmod\ plot in Fig.~\ref{fig:SED_fits} shows that the deviation from the PHOENIX model increases slightly with decreasing frequency below $\nu \sim 400$\,GHz. 
Despite the involved uncertainties, this trend could be interpreted as an indication for a chromospheric temperature rise in  $\gamma$~Lep, as would be expected for an F6V type star. This is also seen in the \Tbp\ in Fig.~\ref{fig:tbprofile}. However, additional observations at frequencies below 230\,GHz (e.g. in band~3) are needed in order to reliably investigate the thermal stratification of a potentially existing chromosphere.  


\paragraph{G --- $\alpha$~Cen~A (HD\,128620J)}  
is of spectral type G2V at a distance of 1.35\,pc with a mass of 1.11\,M$_\odot$ and an age of 4.85\,Gyr. It has a binary companion, $\alpha$~Cen~B, in an orbit with a semi-major axis width of $\asecdot{17}{8}$ and a period of 80~years.
The $\alpha$~Cen~A\&B system was proposed to host a thin disk with a semi-major axis $\approx 4^{\prime\prime}$  based on the observed excess flux density at 24 $\mu$m \citep{Wiegert14}. 
However, the sub-mm/mm observations in ALMA bands 3, 4, 6, 7, 8, and 9 (ALMA observation cycle 2; C2) presented by \citet{2015A&A...573L...4L} and \citet{2016A&A...594A.109L}, and the ATCA observations presented by \citet{2018MNRAS.481..217T} showed no signs of a circumbinary disk.
The ALMA C2 observations had a flux calibration issue that was corrected by \citet{2019arXiv190403043L}, who also presented a new set of observations (Cycle 4; C4) across all the ALMA bands in which the star was previously detected.
The C4 observations performed during 2017 showed that the flux of the star increased since the C2 observations in 2014. 
The resulting time-averaged SED at $\nu < 10^3$\,GHz deviates from the PHOENIX model, with \DSmod\ increasing with decreasing frequency (see Fig.~\ref{fig:SED_fits}).
The bars on the ALMA data points correspond to the variation in the observed flux between C2 and C4 data.
The ATCA flux at 17\,GHz is consistent with this trend, which is indicative of a chromospheric temperature rise in this G-type star \citep{2018MNRAS.481..217T}.  
By comparing and complementing their ALMA measurements with brightness temperatures derived from single-dish observations \citep{2013A&A...549L...7L}, already \citet{2015A&A...573L...4L} and \citet{2016A&A...594A.109L} clearly showed a chromospheric temperature increase and a temperature minimum at a frequency of $\sim 1900$\,GHz ($160\,\mu$m) with a value of $(3920 \pm 375)$\,K \citep{2013A&A...549L...7L}.
This result still stands even with the re-calibrated ALMA data.
Figure~\ref{fig:tbprofile} shows the \Tbp\ in mm -- cm band with re-calibrated ALMA C2 and C4 data marked differently. The C2 data point at 230\,GHz seems to still have some calibration issues, as it is significantly deviant from the C4 value and the general trend.
In general, it is clear that the stellar \Tb\ shows a systematic rise in 2017 C4 observations, provided the re-calibration procedure is now precise.
There is a need for more data with better calibration applied to cross-verify this inference and also to confirm the flux level at 230\,GHz.
However, a hot thermally stratified chromosphere can be inferred, as expected for  this Sun-like star.

\paragraph{H --- 61 Vir (HD\,115617)}  
is a G6.5V star with a mass of 0.92\,M$_\odot$ at a distance of 8.5\,pc. The star has a thin debris disk with a semi-major axis width of $\approx$ 22$^{\prime\prime}$ and an inclination of 77$^\circ$ , as inferred from Spitzer and Herschel data \citep{Bryden06, Wyatt12}. 
It is a multi-planet system with three known planets with estimated masses of 5, 18, and 23 M$_\oplus$ revolving in orbits with semi-major axis widths of 0.05, 0.22, and 0.49 AU, respectively \citep{Vogt10}. The first resolved millimeter ALMA band~7 observations of 61~Vir was presented by \cite{2017MNRAS.469.3518M}. 
The extended disk structure is unseen in ALMA bands, revealing the stellar flux.
Figure ~\ref{fig:SED_fits} presents the SED of the 61 Vir overlaid with the PHOENIX model with an effective temperature of $T_{\rm eff}=5618$\,K.
The excess emission seen in the IR bands above the SED model is likely a result of the disk emission.

\paragraph{I --- $\alpha$~Cen~B (HD\,128621)} 
is the companion to $\alpha$~Cen~A (see above) and has a spectral type of K1V. $\alpha$~Cen~B has an age of 4.85\,Gyr and a mass of 0.90\,M$_\odot$. 
As $\alpha$~Cen~A (see above), $\alpha$~Cen~B has been observed repeatedly with ALMA in almost all so far available receiver bands \citep[3-4 and 6-9; see,][]{2015A&A...573L...4L,2016A&A...594A.109L,2019arXiv190403043L}. 
The observed SED for this star (see Fig.~\ref{fig:SED_fits}) 
shows a deviation from the PHOENIX model SED that increases monotonically towards lower frequencies at $\nu \lesssim 400$\,GHz. This effect can be better seen from \DSmod\ plotted at the bottom of the panel.
Like for its companion, the plotted SED data are the average of ALMA C2 and C4 observations, and the bars on the data points reflect the spread in the values between the two observation epochs.
The corresponding brightness temperatures are presented in Fig.~\ref{fig:tbprofile} and confirm the results of \citet{2016A&A...594A.109L}, who discussed the brightness temperatures in much detail. They concluded that the data for $\alpha$~Cen~B can be matched by a modified solar model that is altered to take into account that $\alpha$~Cen~B is of spectral type K1V. As for $\alpha$~Cen~A, \citet{2015A&A...573L...4L} and \citet{2016A&A...594A.109L} compared and complemented the ALMA measurements with brightness temperatures derived from single-dish observations. The data show a temperature minimum at $\sim 1900$\,GHz with a value of $\sim 3000$\,K for the K-type star $\alpha$~Cen~B, which is thus lower than for the hotter G-type star $\alpha$~Cen~A. Fig.~\ref{fig:tbprofile} shows that the ALMA observations clearly present a chromospheric temperature rise towards values that are indicative of a transition region.

\paragraph{J --- $\epsilon$~Eri (HD\,22049)}  
has a spectral type of K2V and is located at a distance of 3.22\,pc. 
It has a dusty ring with an inner edge around $15^{\prime\prime}$, extending up to $25^{\prime\prime}$ and is well mapped by ALMA \citep{2017MNRAS.469.3200B}. 
Figure ~\ref{fig:SED_fits} presents the SED of the $\epsilon$~Eri overlaid with the PHOENIX model with an effective temperature $T_{\rm eff}=5100$ K.
The VLA and ATCA data provide stellar fluxes between 6 and 45\,GHz \citep{Bastian18_EpsEri, 2019ApJ...871..172R}. 
This star has also been observed in the 217-233\,GHz range using the SMA at an angular resolution of $\asecdot{9}{2}$\,$\times$\,$\asecdot{8}{7}$ \citep{2015ApJ...809...47M}, at 210-290\,GHz using IRAM at an angular resolution of $\asecdot{10}{7}$ \citep{2015A&A...576A..72L}, and at 245-295\,GHz using the LMT at a resolution of $\asecdot{10}{9}$ \citep{ChavezDagostino16}. The fluxes reported by these works are significantly higher than the ALMA value at $\approx$ 230\,GHz, likely due to contamination from the disk emission. 
The flux excess in the range $\nu > 2000$\,GHz to $\nu < 10^4$\,GHz is
also a possible result of contamination from the stellar disk, which is unresolved or barely resolved by the telescopes, as was mentioned in the case of $\alpha$~PsA data. However, ALMA achieves a higher angular resolution of $\asecdot{1}{6}$\,$\times$\,$\asecdot{1}{1}$ and is capable of separating the stellar flux from contributions of the disk. 
At the lowest frequencies, the deviation \DSmod\ increases with decreasing frequency, which implies a chromospheric temperature rise. The available data, however, suggest that a temperature minimum would be mapped at a much lower frequency than what is seen for the Sun, $\alpha$~Cen~A, and $\alpha$~Cen~B.

\paragraph{K --- GJ\,2006A}  
is part of  a common proper motion visual binary  \citep{2014AJ....147...85R,2017ApJ...840...87R} at a distance of 34.8\,pc. Both components are of spectral type  M3.5Ve, with A being brighter than B with respective magnitudes  $B_A=14.33$ and $B_B=14.66$ \citep{2012yCat.1322....0Z}. The two components have a separation of $\asecdot{17}{8}$ \citep{2020ApJS..246....4T}. GJ\,2006A has an age of only 6\,Myr and a mass of 0.55\,\Ms.
We note that for $\nu < 3\times 10^5$\,GHz, the modelled SED from PHOENIX predicts fluxes that are significantly lower than the observed values. This mismatch might be caused by uncertainties in the literature values for the stellar radius, distance, or less likely, the effective temperature and/or by increased flux due to a flare on GJ\,2006A during the time of the observations. 

The star was detected with ALMA in band~7, while  GJ\,2006B remained undetected in that observation. 
From the SED data in Fig.~\ref{fig:SED_fits}, it is clear that the ALMA flux at $\nu = 341$\,GHz is about two orders of magnitude higher than the fluxes predicted by the PHOENIX model. The ALMA flux corresponds to a brightness temperature of $1.6 \times 10^5$\,K, which may indicate the presence of a chromosphere.  
The corresponding excess flux at UV wavelengths seems to support  the existence of a hotter upper atmosphere as well. This interpretation would be consistent with the observed high X-ray luminosity of $\log R_\mathrm{x} = -2.68$ (see Table~\ref{tab:activityprops}). 
However, the excess seems to be much higher than what would be expected from the average (quiescent) state of a chromosphere. Rather, the high values at lower frequencies, the UV excess, and even the mismatch at optical wavelengths might be caused by flares during the individual observations. In addition to potential uncertainties in the stellar parameters and its distance, a consistent increase in observed flux with respect to the modelled SED would require frequent flares, which might be possible for an M3.5Ve star like GJ\,2006A. 
The explanation that the star was in an active phase during the observations would be in line with the results of a long-term VRI photometric study on the GJ 2006AB system, which suggested that either both or one of the stars in the system exhibit long-term variability \citep{2003AJ....125..332J}.  
We conclude that further flux measurements in the mm-cm range are needed to determine the atmospheric temperature stratification and to confirm the existence of a chromosphere for GJ\,2006A and possible flaring. 
It is therefore necessary to establish both the quiescent and active states from repeated observations. In addition, IR observations would allow us to detect or reject possible contributions to a potentially yet unknown disk around this young star. 

\paragraph{L --- Proxima Cen}  
is an M5.5V star that is gravitationally bound to the $\alpha$~Cen system \citep{2017A&A...598L...7K}. It is 
estimated to be 4.85\,Gyr old with a mass of 0.12\,\Ms. It is expected to be a fully convective star, unlike the other stars in the sample, because its mass is below the theoretical core transition threshold of 0.35 \Ms \citep{2000ARA&A..38..337C}. The star was detected with ALMA by \citet{AngladaEscude12}, who proposed the existence of an extended and unresolved disk around the star. However, reanalysis of the same data by \citet{MacGregor18_proxima_flares}  showed that the supposed disk was in fact an imaging artefact. \citeauthor{MacGregor18_proxima_flares}   also reported the discovery of a short-duration flare that lasted for a few minutes. 
The pre-flare flux was about 20\% higher than the value derived from the photospheric SED modelling  based on  optical to IR fluxes by \citet{Ribas17_ProximaCenSED}.
This slight excess in flux is most likely due to a hot chromosphere. 
Another flare observed in 2019 lasted for a few seconds in ALMA band 6, along with simultaneous flaring in the  optical to far-UV bands \citep{2021ApJ...911L..25M}.

Moreover, the reduction of the SED fluxes in the 10$^5$ -- 10$^6$\,GHz range is noteworthy. This is unlike the case for GJ\,2006\,A, which is a much younger M~dwarf, although it is more massive and expected to be less active than a fully convective star such as Proxima Cen.


\textbf{}
\section{Discussion}
\label{sec:discussion}

\subsection{Thermal versus non-thermal emission in the mm-cm range}
\label{sec:emission_in_mm-cm}

Observations of the Sun at mm wavelengths have  testified that the emission is optically thick thermal radiation from various atmospheric layers with brightness temperatures (\Tb) of 21000\,K, 9000\,K, 7300\,K, and 5900\,K at 34\,GHz, 85\,GHz, 93\,GHz, and 233\,GHz, respectively \citep{2006A&A...456..697W, 2008A&A...488.1079S, 2017SoPh..292...88W}. 
The formation height of the continuum radiation in the solar atmosphere increases with decreasing frequency in the mm range \citep[see e.g.][and references therein]{2016SSRv..200....1W}, which allows determining the temperature stratification from measurements as listed above. 
The effect is also clearly seen from the detailed contribution functions for the solar emission in the entire mm -- cm band based on the 1D atmospheric model by \citet{2008ApJS..175..229A} as presented by  \citet{2011SoPh..273..309S}. We note that the contribution functions by \citet{2011SoPh..273..309S} can span rather extended height ranges in the mm range, which is in line with results from 3D simulations of the solar atmosphere \citep[e.g.][]{Loukitcheva15_mmRad_model}. However, it should be mentioned that the precise formation heights of the solar continuum radiation at millimeter wavelengths are still debated
\citep{2016SSRv..200....1W}. 
For the 10 -- 30\,GHz range, that is, at cm wavelengths, the contribution functions by \citet{2011SoPh..273..309S} show that the radiation originates from a compressed height range in the upper chromosphere and transition region. 
Observations of the Sun in a quiescent state at 10\,GHz revealed a mean \Tb\ of 20,000\,K that is typically seen in the transition region \citep{1991ApJ...370..779Z}. 
During strong flares, however, gyro-resonance and gyro-synchrotron emission from supra-thermal electrons has been observed close to the 10\,GHz  range  in coronal active regions \citep[e.g.][]{Nindos00_SXR_gyroresonance_Sunspots, 2006PASJ...58...11V}.
The highest frequency at which gyro-resonance emission has been conclusively detected so far from solar active regions is 17\,GHz \citep{2020FrASS...7...57N}. 
In this regard, it should be noted that gyro-resonance or gyro-synchrotron emission sources  are observed less frequently at $\nu > 10$\,GHz than in the $\nu \sim$ 1 -- 5\,GHz range, which is more sensitive to such emission.
Analysing the number of active regions detected in daily solar images at 17\,GHz archived during 1993 -- 2013 by the Nobeyama Radio Heliograph, \citet{Selhorst14_NoRH_SSN_3GHz_srcCount} showed that the annual mean counts of active regions during that period lie in the range  $\approx$ 0 -- 4. Even during the period of maximum solar activity, the annual mean counts were just within 3 -- 4, when the average annual counts of sunspots and gyro-resonance sources at $\approx$ 3\,GHz varied around 100 -- 130 and 160 -- 180, respectively.
A physical reason behind this dearth of sources is that the gyro-frequency ($\nu_B$) for a given magnetic field strength, $B$, is given by
\begin{equation}
    \nu_\mathrm{B} = 2.9\times \left( \frac{B}{1\,G}\right)\, \mathrm{MHz}.
    \label{Eqn:nuB}
\end{equation}

Typically, the gyro-resonance emission is observed up to the third$^{}$ harmonic of $\nu_\mathrm{B}$ \citep{1997SoPh..174...31W}. Consequently, gyro-resonance emission beyond 10\,GHz can only be detected for coronal magnetic field strengths of at least $\approx$ 1150\,G. Typical magnetic field strengths at coronal active regions, however, range within $\approx$ 100 -- 500\,G \citep{1978SoPh...57..279D, 1999ApJ...515..842A}. Unusually large field strengths over 1000\,G have been reported occasionally in flaring coronal active regions near sunspots with photospheric field strengths close to 2500\,G \citep[e.g.][]{1991ApJ...366L..43W,1994SSRv...68..217S}, however.
We therefore conclude that in the Sun, non-thermal contributions above 10\,GHz are expected only in extreme situations such as flares. The situation might differ in other stars depending on their magnetic field strengths and topology and on the resulting level of activity.  
For low-mass M dwarf stars with stronger photospheric magnetic field concentrations in the 2000 -- 5000\,G range,   \citet{2010MNRAS.407.2269M} could show gyro-synchrotron emission even above 20\,GHz in the upper atmospheric layers, depending on which harmonic is observed. 
Based on archival observations of M dwarfs in microwave bands ($\nu < 10$\,GHz), it has been concluded that they show signatures of gyro-synchrotron emission even during their quiescent phase up to about 20\,GHz \citep{1990SoPh..130..265B}.
From this discussion, we conclude that to model the mm -- cm \Tbp\ of  low-mass M dwarfs, non-thermal contributions cannot be neglected even for the emission during non-flaring times. In the case of Sun-like stars, it is important to observe the star for a longer period so that a light curve can be obtained and any occasional high-energy non-thermal contributions from flares can be systematically filtered out to robustly detect the quiescent thermal emission.
This would then mean that any inference about the emission mechanism behind the detected fluxes from Sun-like stars close 10\,GHz should be made with caution unless the temporal variation and occurrence of flares can be deduced from an observed light curve.

\subsection{Dependence on spectral type} 
\label{sec:TB_profile}

This section focuses on the derived spectral flux excess (\DSmod) for stars with a sufficient amount of data across a wider spectral range. Stellar \DSmod\ spectra are compared against that for the Sun as a star.
As defined in Sect.~\ref{sec:fluxexcessparameter}, \DSmod, describes the deviations of the observed fluxes from those predicted by a model that features a photosphere, but lacks a chromosphere with a notable temperature rise. This parameter could also be interpreted as a proxy for the additional 
 thermal energy content in the upper stellar atmospheric layers that is connected to the presence of a stellar chromosphere. 
A positive (negative) deviation in the \Snu, from the PHOENIX model prediction, \Smod, could translate into an additional heating (cooling) or emission (absorption) mechanism operating in those atmospheric layers. These emission sources could be inherent to the stellar atmosphere or dissociated from it, but spatially unresolved by the telescope, such as disks or companions. The \DSmod\ spectrum can give indications about the nature of this additional mechanism or source. For instance, at frequencies close to 10\,GHz, non-thermal gyro-synchrotron emission from energetic particles in active regions could contribute significantly to \Snu, resulting in a positive \DSmod\ for  FGKM stars. 
\DSmod\ can thus be a proxy to the excess combined thermal and non-thermal energy content across the upper atmospheric layers. 
However, to determine the fractional contribution from thermal emission and gyro-synchrotron emission to \Snu, a detailed atmospheric emission modelling with input from magnetic field measurements for the star is needed. 
We would also require stellar flux data sampled more densely and uniformly in the mm -- cm range because the stellar upper atmospheric layers for Sun-like stars are expected to have strong gradients in their thermal and magnetic stratification 
\citep[e.g.][]{1981ApJS...45..635V,gary2001,2013A&A...549L...7L}.
The \DSmod\ spectrum derived from such a well-sampled spectrum can demonstrate the changes in the dominant emission mechanism at different atmospheric layers probed because these changes often cause deviations in the spectral shape.
In this case,   \DSmod\ in the $10 - 100$\,GHz range for M dwarfs should show signs of deviations from its high-frequency spectral structure owing to the expected significance of gyro-synchrotron emission.

Using Eqn.~\ref{Eqn:DelS}, 
we derived the spectral flux excess for the stars in the sample, ranging from 0.5 - 2.1\,\Ms (A1 - M3.5). The \DSmod\ spectra for stars are provided in the bottom panel of the SED plots in Fig.~\ref{fig:SED_fits}.
The \DSmod\ values fluctuate around zero for the A-type star $\alpha$\,CMa. For the F6 star $\gamma$\,Lep, however, we find that \DSmod\ shows a tendency to rise with decreasing frequency. This rising trend is even clearer for the G-type stars, the Sun and $\alpha$\,Cen\,A, owing to the better spectral sampling of data.
It is notable that the rate of increase of \DSmod\ with decreasing frequency, becomes steeper for stars with lower \teff. 
A similar effect is seen in the \Tbp\ of G -- K dwarfs in Fig.~\ref{fig:tbprofile} as well.

Figure~\ref{fig:DS_by_S_spectrum} shows the variation of \DSmod\ as a function of \teff\ for the stars in the sample for ALMA bands 7 and 6. 
\DSmod\ shows a sign of an increasing trend with decreasing \teff\ in band~6 for stars with \teff\ within 4000 -- 10000\,K. The single fully convective cool star, Proxima\,Cen, showed a \DSmod\ value lower than the K-type star $\epsilon$~Eri, deviating from this trend.
For the hottest star, $\alpha$\,CMa~(A), \DSmod$\approx 0$ in band~6, which means that the observed flux is well reproduced by the corresponding PHOENIX model without a chromosphere. 
Because the stellar sample was chosen to have no contamination from any companion or disk in the mm--cm band range, the excess flux can be attributed to the excess thermal and/or non-thermal energy content that is connected to the presence of a stellar chromosphere.
All the stars except for those with ID A and B belong to F, G, K, or M spectral type, which are expected to have a hot chromosphere. 
The \DSmod\ values of C and D ($\gamma$~Vir\,A\&B) appear to significantly deviate from the general trend, but are still in line with the trend within the considerable error bars.
In the plot for band 7, we find that all the FGK stars show a \DSmod\ $\sim$ 0 -- 1, whereas the M dwarf, GJ\,2006\,A (ID: K), has a significantly higher value (12000\%) in its  atmospheric layer probed by band 7. This could be the result of additional non-thermal contributions or even a flare. As already mentioned in Sect.~\ref{sec:stellarsample}, this explanation is plausible because M dwarfs are known to be frequently flaring \citep{Lacy76_Flare_stats_Mstars, 2020MNRAS.494.4848D}. With just one \Tb\ point in the whole mm -- cm band, however, it is hard to draw any robust conclusion. 
On the other hand, it is quite intriguing that the \DSmod\ of Proxima\,Cen is much lower than for GJ\,2006\,A, even though it belongs to a class of more active fully convective stars.
They have much stronger and topologically quite different magnetic field structure than other stars \citep[e.g]{Donati09_Rev_Bfield,Lund20_stelar_helicity}.
By virtue of this, Proxima\,Cen could possess a very different atmospheric structure, emission mechanism and optical depth at different spectral bands compared to GJ\,2006\,A.
Light curves at different mm -- cm wavelengths are required to better understand the emission characteristics of these M dwarfs. This would help us to systematically isolate flaring and non-flaring fluxes in the time domain  and study their quiescent spectra. Performing such a study across FGKM dwarfs of masses distributed around the theoretical core transition threshold of 0.35\Ms\  can help understand the differences in their atmospheric stratification, if any, as a result of the internal transition.
In addition to mass, age is another crucial factor that determines the physical structure and properties of a star \citep{Barnes03_Rot_Vs_age_Vs_Activity,Lund20_stelar_helicity}. The fact that Proxima\,Cen is 4.8\,Gyr old while GJ\,2006\,A is just 6\,Myr old could play an important role in the observed large difference in their \DSmod\ values.  

This study demonstrates the importance of carrying out a systematic \DSmod\ spectral evolution study for stars belonging to different types and ages. 
The mm -- cm band could help us understand how the  atmospheric excess energy budget and corresponding atmospheric heating varies for these stars as function of spectral type and age.


\subsection{Possible extension of the sample size}
\label{sec:disc_samplesize}

The sample presented in this study is intended to be a starting point to illustrate the potential scientific value of a larger sample. 
To understand the possibilities of extending the sample, it is import to have an estimate of the largest distance up to which ALMA can detect a main-sequence star. 
Consider an observation at $240$\,GHz($\approx 1.25\,$mm), which lies within band~6, one of the most sensitive and widely used ALMA receiver bands. An observation in this band using all available antennas of ALMA  with full spectral averaging and an integration time of 1 hour can provide a $3 \sigma$ source detection threshold of around $50\,\mu$Jy. 
Assuming a typical temperature of $5900$\,K for the Sun in band 6 \citep{2017SoPh..292...88W}, with the aforementioned detection threshold ALMA will detect a Sun-like star located within $10$\,pc. The recent most extensive catalogue of stars, Gaia DR2 (\citealp{Gaia18_DR2_catalog}), reports about $1675$ potential main-sequence stars of type F, G, K in this volume.
In addition, because F stars are hotter and because KM stars can have higher  \Tb\ than the Sun in band~6 owing to their relatively steeper \Tbp\ compared to  G-type stars in the $\sim$ 10 -- 300\,GHz band (see, Fig.~\ref{fig:tbprofile}), they could be detected at distances even farther out than 10\,pc. 


\section{Conclusions and outlook}
\label{sec:conclus}

The emission of main-sequence stars in the millimeter to centimeter wavelength range (10 -- 1000\,GHz)  originates from their outer atmospheric layers, such that the height of the emitting layer increases with decreasing frequency. In the case of Sun-like stars with an outer convective zone, the emission is expected to originate from the  chromosphere and corona.
This work analysed the features of 10 -- 1000\,GHz spectra of main-sequence stars with the aim to characterise the corresponding thermal stratification in their atmospheres including the presence of a chromospheric temperature rise and potential non-thermal contributions at the lowest frequencies.  
A search of archives and the literature for significant detections of main-sequence stars with ALMA resulted in a sample of 12~stars across a broad range of spectral types from A1 to M3.5. 
%
The centimeter band spectra of these stars were also compiled to obtain the spectrum in the 10 -- 1000\,GHz range.
We then defined a spectral flux excess parameter, \DSmod, as the ratio of the difference between the observed and the PHOENIX model spectral flux density to the respective model values. 
Because the stars were chosen so that no known disks, companions, or other emission sources contaminated the detected stellar emission at mm -- cm wavelengths, the resulting \DSmod\ spectrum in the mm -- cm range is purely star-borne. We find a systematic rise in the \DSmod\ of the FGKM stars with decreasing observation frequency. For the stars within $T_\mathrm{eff} = 4000 - 10000$\,K, the rate of the rise becomes steeper as the effective temperature (\teff) of the star decreases. 

The \DSmod\ values as function of \teff\ were studied at two representative ALMA bands (bands 6 and 7). 
The A-type stars were found to show \DSmod\ $\sim$ 0 in both bands, whereas the FGKM stars showed a systematic spectral flux excess. This demonstrates that the spectral flux densities of A-type stars can be matched with models without a chromosphere, as expected for this stellar type. The positive \DSmod\ for cooler stars implies significant outer atmospheric heating and thus the existence of a hot chromosphere. 
For the single fully convective star in our sample,  \DSmod\ was greater than 0, although the value was significantly lower than what would be expected from the apparent trend for the FGKM stars. 
In band 7, GJ\,2006\,A showed a flux excess of $\sim$ 1100\%. With data available for this star in only one time and frequency point in the mm -- cm band, it is hard to draw any inferences about the nature of the energy source that powered this emission. It is also possible that this flux was associated with a transient flare because M dwarfs are known to be frequently flaring.

This study thus demonstrated the unique potential of mm -- cm band long-term observations of main-sequence stars to better understand the energy budget and temperature across their different outer atmospheric layers. Such periodic observation campaigns over years can help understand the nexus between the stellar activity cycles or variations and the atmospheric structure. By analysing the observations using detailed thermal and non-thermal emission models, the physical state of the plasma, its dynamics, and the nature of activity can be better understood.

\section*{Acknowledgments}
This work is supported by the Research Council of Norway through its Centres of Excellence scheme, project number 262622 (``Rosseland Centre for Solar Physics'').  
AM and SP acknowledge support from the EMISSA project funded by the Research Council of Norway (project number 286853).
SW and MS were supported  by the SolarALMA project, which has received funding from the European Research Council (ERC) under the European Union’s Horizon 2020 research and innovation programme (grant agreement No. 682462).  
This paper makes use of the following ALMA data: ADS/JAO.ALMA\#2012.1.00238.S, ADS/JAO.ALMA\#2012.1.00385.S, ADS/JAO.ALMA\#2013.1.00170.S, ADS/JAO.ALMA\#2013.1.00359.S, ADS/JAO.ALMA\#2013.1.00645.S, ADS/JAO.ALMA\#2016.A.00013.S, ADS/JAO.ALMA\#2016.1.00441.S, ADS/JAO.ALMA\#2017.1.00698.S, ADS/JAO.ALMA\#2017.1.01583.S, ADS/JAO.ALMA\#2018.1.00470.S, ADS/JAO.ALMA\#2018.1.01149.S.
ALMA is a partnership of ESO (representing its member states), NSF (USA) and NINS (Japan), together with NRC(Canada), MOST and ASIAA (Taiwan), and KASI (Republic of Korea), in co-operation with the Republic of Chile. The Joint ALMA Observatory is operated by ESO, AUI/NRAO and NAOJ. 
SW and PHH acknowledge fruitful discussions 
of the International Team 265 (\textit{Magnetic Activity of M-type Dwarf Stars and the Influence on Habitable Extra-Solar Planets}) supported by the International Space Science Institute, Bern, Switzerland. 
Some of the  calculations presented here were performed at the RRZ
of the Universit\"at Hamburg, at the H\"ochstleistungs Rechenzentrum Nord
(HLRN), and at the National Energy Research Supercomputer Center (NERSC), which
is supported by the Office of Science of the U.S.  Department of Energy under
Contract No. DE-AC03-76SF00098.  We thank all these institutions for a generous
allocation of computer time.  PHH gratefully acknowledges the support of NVIDIA
Corporation with the donation of a Quadro P6000 GPU used in this research.


\bibliographystyle{aa}
\bibliography{EMISSA_allref} 

\begin{thebibliography}{162}
\expandafter\ifx\csname natexlab\endcsname\relax\def\natexlab#1{#1}\fi

\bibitem[{{Acke} {et~al.}(2012){Acke}, {Min}, {Dominik}, {Vandenbussche},
  {Sibthorpe}, {Waelkens}, {Olofsson}, {Degroote}, {Smolders}, {Pantin},
  {Barlow}, {Blommaert}, {Brandeker}, {De Meester}, {Dent}, {Exter}, {Di
  Francesco}, {Fridlund}, {Gear}, {Glauser}, {Greaves}, {Harvey}, {Henning},
  {Hogerheijde}, {Holland}, {Huygen}, {Ivison}, {Jean}, {Liseau}, {Naylor},
  {Pilbratt}, {Polehampton}, {Regibo}, {Royer}, {Sicilia-Aguilar}, \&
  {Swinyard}}]{2012A&A...540A.125A}
{Acke}, B., {Min}, M., {Dominik}, C., {et~al.} 2012, \aap, 540, A125

\bibitem[{{Akeson} {et~al.}(2021){Akeson}, {Beichman}, {Kervella}, {Fomalont},
  \& {Benedict}}]{2021AJ....162...14A}
{Akeson}, R., {Beichman}, C., {Kervella}, P., {Fomalont}, E., \& {Benedict},
  G.~F. 2021, \aj, 162, 14

\bibitem[{{Allende Prieto} \& {Lambert}(1999)}]{Prieto99_Ref4}
{Allende Prieto}, C. \& {Lambert}, D.~L. 1999, \aap, 352, 555

\bibitem[{{Ammler-von Eiff} \& {Reiners}(2012)}]{2012A&A...542A.116A}
{Ammler-von Eiff}, M. \& {Reiners}, A. 2012, \aap, 542, A116

\bibitem[{{Anglada-Escud{\'e}} \& {Butler}(2012)}]{AngladaEscude12}
{Anglada-Escud{\'e}}, G. \& {Butler}, R.~P. 2012, \apjs, 200, 15

\bibitem[{{Anstee} {et~al.}(1997){Anstee}, {O'Mara}, \&
  {Ross}}]{1997MNRAS.284..202A}
{Anstee}, S.~D., {O'Mara}, B.~J., \& {Ross}, J.~E. 1997, \mnras, 284, 202

\bibitem[{{Aschwanden} {et~al.}(1999){Aschwanden}, {Newmark},
  {Delaboudini{\`e}re}, {Neupert}, {Klimchuk}, {Gary}, {Portier-Fozzani}, \&
  {Zucker}}]{1999ApJ...515..842A}
{Aschwanden}, M.~J., {Newmark}, J.~S., {Delaboudini{\`e}re}, J.-P., {et~al.}
  1999, \apj, 515, 842

\bibitem[{{Avrett} \& {Loeser}(2008)}]{2008ApJS..175..229A}
{Avrett}, E.~H. \& {Loeser}, R. 2008, \apjs, 175, 229

\bibitem[{Barnes(2003)}]{Barnes03_Rot_Vs_age_Vs_Activity}
Barnes, S.~A. 2003, The Astrophysical Journal, 586, 464

\bibitem[{{Bastian}(1990)}]{1990SoPh..130..265B}
{Bastian}, T.~S. 1990, \solphys, 130, 265

\bibitem[{{Bastian}(1994)}]{1994SSRv...68..261B}
{Bastian}, T.~S. 1994, \ssr, 68, 261

\bibitem[{{Bastian} {et~al.}(2018){Bastian}, {Villadsen}, {Maps}, {Hallinan},
  \& {Beasley}}]{Bastian18_EpsEri}
{Bastian}, T.~S., {Villadsen}, J., {Maps}, A., {Hallinan}, G., \& {Beasley},
  A.~J. 2018, \apj, 857, 133

\bibitem[{{Bazot} {et~al.}(2007){Bazot}, {Bouchy}, {Kjeldsen}, {Charpinet},
  {Laymand}, \& {Vauclair}}]{2007A&A...470..295B}
{Bazot}, M., {Bouchy}, F., {Kjeldsen}, H., {et~al.} 2007, \aap, 470, 295

\bibitem[{{Bond} {et~al.}(2017){Bond}, {Schaefer}, {Gilliland}, {Holberg},
  {Mason}, {Lindenblad}, {Seitz-McLeese}, {Arnett}, {Demarque}, {Spada},
  {Young}, {Barstow}, {Burleigh}, \& {Gudehus}}]{2017ApJ...840...70B}
{Bond}, H.~E., {Schaefer}, G.~H., {Gilliland}, R.~L., {et~al.} 2017, \apj, 840,
  70

\bibitem[{{Booth} {et~al.}(2017){Booth}, {Dent}, {Jord{\'a}n}, {Lestrade},
  {Hales}, {Wyatt}, {Casassus}, {Ertel}, {Greaves}, {Kennedy}, {Matr{\`a}},
  {Augereau}, \& {Villard}}]{2017MNRAS.469.3200B}
{Booth}, M., {Dent}, W. R.~F., {Jord{\'a}n}, A., {et~al.} 2017, \mnras, 469,
  3200

\bibitem[{{Boro Saikia} {et~al.}(2018){Boro Saikia}, {Marvin}, {Jeffers},
  {Reiners}, {Cameron}, {Marsden}, {Petit}, {Warnecke}, \&
  {Yadav}}]{2018A&A...616A.108B}
{Boro Saikia}, S., {Marvin}, C.~J., {Jeffers}, S.~V., {et~al.} 2018, \aap, 616,
  A108

\bibitem[{{Bryden} {et~al.}(2006){Bryden}, {Beichman}, {Trilling}, {Rieke},
  {Holmes}, {Lawler}, {Stapelfeldt}, {Werner}, {Gautier}, {Blaylock}, {Gordon},
  {Stansberry}, \& {Su}}]{Bryden06}
{Bryden}, G., {Beichman}, C.~A., {Trilling}, D.~E., {et~al.} 2006, \apj, 636,
  1098

\bibitem[{{Bychkov} {et~al.}(2009){Bychkov}, {Bychkova}, \&
  {Madej}}]{2009MNRAS.394.1338B}
{Bychkov}, V.~D., {Bychkova}, L.~V., \& {Madej}, J. 2009, \mnras, 394, 1338

\bibitem[{{Cayrel de Strobel} {et~al.}(2001){Cayrel de Strobel}, {Soubiran}, \&
  {Ralite}}]{2001A&A...373..159C}
{Cayrel de Strobel}, G., {Soubiran}, C., \& {Ralite}, N. 2001, \aap, 373, 159

\bibitem[{{Chabrier} \& {Baraffe}(2000)}]{2000ARA&A..38..337C}
{Chabrier}, G. \& {Baraffe}, I. 2000, \araa, 38, 337

\bibitem[{{Chavez-Dagostino} {et~al.}(2016){Chavez-Dagostino}, {Bertone},
  {Cruz-Saenz de Miera}, {Marshall}, {Wilson}, {S{\'a}nchez-Arg{\"u}elles},
  {Hughes}, {Kennedy}, {Vega}, {De la Luz}, {Dent}, {Eiroa}, {G{\'o}mez-Ruiz},
  {Greaves}, {Lizano}, {L{\'o}pez-Valdivia}, {Mamajek}, {Monta{\~n}a},
  {Olmedo}, {Rodr{\'\i}guez-Montoya}, {Schloerb}, {Yun}, {Zavala}, \&
  {Zeballos}}]{ChavezDagostino16}
{Chavez-Dagostino}, M., {Bertone}, E., {Cruz-Saenz de Miera}, F., {et~al.}
  2016, \mnras, 462, 2285

\bibitem[{{Coddington} {et~al.}(2019){Coddington}, {Lean}, {Pilewskie}, {Snow},
  {Richard}, {Kopp}, {Lindholm}, {DeLand}, {Marchenko}, {Haberreiter}, \&
  {Baranyi}}]{2019E&SS....6.2525C}
{Coddington}, O., {Lean}, J., {Pilewskie}, P., {et~al.} 2019, Earth and Space
  Science, 6, 2525

\bibitem[{{Crosley} \& {Osten}(2018)}]{2018ApJ...856...39C}
{Crosley}, M.~K. \& {Osten}, R.~A. 2018, \apj, 856, 39

\bibitem[{{David} \& {Hillenbrand}(2015)}]{2015ApJ...804..146D}
{David}, T.~J. \& {Hillenbrand}, L.~A. 2015, \apj, 804, 146

\bibitem[{{Davis} {et~al.}(2020){Davis}, {Taylor}, \&
  {Dowell}}]{2020MNRAS.494.4848D}
{Davis}, I., {Taylor}, G., \& {Dowell}, J. 2020, \mnras, 494, 4848

\bibitem[{{DeWarf} {et~al.}(2010){DeWarf}, {Datin}, \&
  {Guinan}}]{2010ApJ...722..343D}
{DeWarf}, L.~E., {Datin}, K.~M., \& {Guinan}, E.~F. 2010, \apj, 722, 343

\bibitem[{{Di Folco} {et~al.}(2004){Di Folco}, {Th{\'e}venin}, {Kervella},
  {Domiciano de Souza}, {Coud{\'e} du Foresto}, {S{\'e}gransan}, \&
  {Morel}}]{DiFolco04}
{Di Folco}, E., {Th{\'e}venin}, F., {Kervella}, P., {et~al.} 2004, \aap, 426,
  601

\bibitem[{{Donati} \& {Landstreet}(2009)}]{Donati09_Rev_Bfield}
{Donati}, J.~F. \& {Landstreet}, J.~D. 2009, \araa, 47, 333

\bibitem[{{Duch{\^e}ne} {et~al.}(2014){Duch{\^e}ne}, {Arriaga}, {Wyatt},
  {Kennedy}, {Sibthorpe}, {Lisse}, {Holland}, {Wisniewski}, {Clampin}, {Kalas},
  {Pinte}, {Wilner}, {Booth}, {Horner}, {Matthews}, \&
  {Greaves}}]{2014ApJ...784..148D}
{Duch{\^e}ne}, G., {Arriaga}, P., {Wyatt}, M., {et~al.} 2014, \apj, 784, 148

\bibitem[{{Dulk} \& {McLean}(1978)}]{1978SoPh...57..279D}
{Dulk}, G.~A. \& {McLean}, D.~J. 1978, \solphys, 57, 279

\bibitem[{{Fabricius} {et~al.}(2002){Fabricius}, {H{\o}g}, {Makarov}, {Mason},
  {Wycoff}, \& {Urban}}]{2002A&A...384..180F}
{Fabricius}, C., {H{\o}g}, E., {Makarov}, V.~V., {et~al.} 2002, \aap, 384, 180

\bibitem[{{Fazio} {et~al.}(2004){Fazio}, {Hora}, {Allen}, {Ashby}, {Barmby},
  {Deutsch}, {Huang}, {Kleiner}, {Marengo}, {Megeath}, {Melnick}, {Pahre},
  {Patten}, {Polizotti}, {Smith}, {Taylor}, {Wang}, {Willner}, {Hoffmann},
  {Pipher}, {Forrest}, {McMurty}, {McCreight}, {McKelvey}, {McMurray}, {Koch},
  {Moseley}, {Arendt}, {Mentzell}, {Marx}, {Losch}, {Mayman}, {Eichhorn},
  {Krebs}, {Jhabvala}, {Gezari}, {Fixsen}, {Flores}, {Shakoorzadeh}, {Jungo},
  {Hakun}, {Workman}, {Karpati}, {Kichak}, {Whitley}, {Mann}, {Tollestrup},
  {Eisenhardt}, {Stern}, {Gorjian}, {Bhattacharya}, {Carey}, {Nelson},
  {Glaccum}, {Lacy}, {Lowrance}, {Laine}, {Reach}, {Stauffer}, {Surace},
  {Wilson}, {Wright}, {Hoffman}, {Domingo}, \& {Cohen}}]{2004ApJS..154...10F}
{Fazio}, G.~G., {Hora}, J.~L., {Allen}, L.~E., {et~al.} 2004, \apjs, 154, 10

\bibitem[{{Feinstein} {et~al.}(2020){Feinstein}, {Montet}, {Ansdell}, {Nord},
  {Bean}, {G{\"u}nther}, {Gully-Santiago}, \&
  {Schlieder}}]{2020AJ....160..219F}
{Feinstein}, A.~D., {Montet}, B.~T., {Ansdell}, M., {et~al.} 2020, \aj, 160,
  219

\bibitem[{{Freund} {et~al.}(2018){Freund}, {Robrade}, {Schneider}, \&
  {Schmitt}}]{2018A&A...614A.125F}
{Freund}, S., {Robrade}, J., {Schneider}, P.~C., \& {Schmitt}, J.~H.~M.~M.
  2018, \aap, 614, A125

\bibitem[{{Fr{\"o}hlich}(2007)}]{2007AN....328.1037F}
{Fr{\"o}hlich}, H.~E. 2007, Astronomische Nachrichten, 328, 1037

\bibitem[{{Fuhrmann} {et~al.}(2017){Fuhrmann}, {Chini}, {Kaderhandt}, \&
  {Chen}}]{2017ApJ...836..139F}
{Fuhrmann}, K., {Chini}, R., {Kaderhandt}, L., \& {Chen}, Z. 2017, \apj, 836,
  139

\bibitem[{{Gaia Collaboration} {et~al.}(2018){Gaia Collaboration}, {Brown},
  {Vallenari}, {Prusti}, {de Bruijne}, {Babusiaux}, {Bailer-Jones}, {Biermann},
  {Evans}, {Eyer}, {Jansen}, {Jordi}, {Klioner}, {Lammers}, {Lindegren},
  {Luri}, {Mignard}, {Panem}, {Pourbaix}, {Randich}, {Sartoretti}, {Siddiqui},
  {Soubiran}, {van Leeuwen}, {Walton}, {Arenou}, {Bastian}, {Cropper},
  {Drimmel}, {Katz}, {Lattanzi}, {Bakker}, {Cacciari}, {Casta{\~n}eda},
  {Chaoul}, {Cheek}, {De Angeli}, {Fabricius}, {Guerra}, {Holl}, {Masana},
  {Messineo}, {Mowlavi}, {Nienartowicz}, {Panuzzo}, {Portell}, {Riello},
  {Seabroke}, {Tanga}, {Th{\'e}venin}, {Gracia-Abril}, {Comoretto},
  {Garcia-Reinaldos}, {Teyssier}, {Altmann}, {Andrae}, {Audard},
  {Bellas-Velidis}, {Benson}, {Berthier}, {Blomme}, {Burgess}, {Busso},
  {Carry}, {Cellino}, {Clementini}, {Clotet}, {Creevey}, {Davidson}, {De
  Ridder}, {Delchambre}, {Dell'Oro}, {Ducourant},
  {Fern{\'a}ndez-Hern{\'a}ndez}, {Fouesneau}, {Fr{\'e}mat}, {Galluccio},
  {Garc{\'\i}a-Torres}, {Gonz{\'a}lez-N{\'u}{\~n}ez}, {Gonz{\'a}lez-Vidal},
  {Gosset}, {Guy}, {Halbwachs}, {Hambly}, {Harrison}, {Hern{\'a}ndez},
  {Hestroffer}, {Hodgkin}, {Hutton}, {Jasniewicz}, {Jean-Antoine-Piccolo},
  {Jordan}, {Korn}, {Krone-Martins}, {Lanzafame}, {Lebzelter}, {L{\"o}ffler},
  {Manteiga}, {Marrese}, {Mart{\'\i}n-Fleitas}, {Moitinho}, {Mora}, {Muinonen},
  {Osinde}, {Pancino}, {Pauwels}, {Petit}, {Recio-Blanco}, {Richards},
  {Rimoldini}, {Robin}, {Sarro}, {Siopis}, {Smith}, {Sozzetti}, {S{\"u}veges},
  {Torra}, {van Reeven}, {Abbas}, {Abreu Aramburu}, {Accart}, {Aerts},
  {Altavilla}, {{\'A}lvarez}, {Alvarez}, {Alves}, {Anderson}, {Andrei},
  {Anglada Varela}, {Antiche}, {Antoja}, {Arcay}, {Astraatmadja}, {Bach},
  {Baker}, {Balaguer-N{\'u}{\~n}ez}, {Balm}, {Barache}, {Barata}, {Barbato},
  {Barblan}, {Barklem}, {Barrado}, {Barros}, {Barstow}, {Bartholom{\'e}
  Mu{\~n}oz}, {Bassilana}, {Becciani}, {Bellazzini}, {Berihuete}, {Bertone},
  {Bianchi}, {Bienaym{\'e}}, {Blanco-Cuaresma}, {Boch}, {Boeche}, {Bombrun},
  {Borrachero}, {Bossini}, {Bouquillon}, {Bourda}, {Bragaglia}, {Bramante},
  {Breddels}, {Bressan}, {Brouillet}, {Br{\"u}semeister}, {Brugaletta},
  {Bucciarelli}, {Burlacu}, {Busonero}, {Butkevich}, {Buzzi}, {Caffau},
  {Cancelliere}, {Cannizzaro}, {Cantat-Gaudin}, {Carballo}, {Carlucci},
  {Carrasco}, {Casamiquela}, {Castellani}, {Castro-Ginard}, {Charlot},
  {Chemin}, {Chiavassa}, {Cocozza}, {Costigan}, {Cowell}, {Crifo}, {Crosta},
  {Crowley}, {Cuypers}, {Dafonte}, {Damerdji}, {Dapergolas}, {David}, {David},
  {de Laverny}, {De Luise}, {De March}, {de Martino}, {de Souza}, {de Torres},
  {Debosscher}, {del Pozo}, {Delbo}, {Delgado}, {Delgado}, {Di Matteo},
  {Diakite}, {Diener}, {Distefano}, {Dolding}, {Drazinos}, {Dur{\'a}n},
  {Edvardsson}, {Enke}, {Eriksson}, {Esquej}, {Eynard Bontemps}, {Fabre},
  {Fabrizio}, {Faigler}, {Falc{\~a}o}, {Farr{\`a}s Casas}, {Federici},
  {Fedorets}, {Fernique}, {Figueras}, {Filippi}, {Findeisen}, {Fonti},
  {Fraile}, {Fraser}, {Fr{\'e}zouls}, {Gai}, {Galleti}, {Garabato},
  {Garc{\'\i}a-Sedano}, {Garofalo}, {Garralda}, {Gavel}, {Gavras}, {Gerssen},
  {Geyer}, {Giacobbe}, {Gilmore}, {Girona}, {Giuffrida}, {Glass}, {Gomes},
  {Granvik}, {Gueguen}, {Guerrier}, {Guiraud}, {Guti{\'e}rrez-S{\'a}nchez},
  {Haigron}, {Hatzidimitriou}, {Hauser}, {Haywood}, {Heiter}, {Helmi}, {Heu},
  {Hilger}, {Hobbs}, {Hofmann}, {Holland}, {Huckle}, {Hypki}, {Icardi},
  {Jan{\ss}en}, {Jevardat de Fombelle}, {Jonker}, {Juh{\'a}sz}, {Julbe},
  {Karampelas}, {Kewley}, {Klar}, {Kochoska}, {Kohley}, {Kolenberg},
  {Kontizas}, {Kontizas}, {Koposov}, {Kordopatis}, {Kostrzewa-Rutkowska},
  {Koubsky}, {Lambert}, {Lanza}, {Lasne}, {Lavigne}, {Le Fustec}, {Le
  Poncin-Lafitte}, {Lebreton}, {Leccia}, {Leclerc}, {Lecoeur-Taibi},
  {Lenhardt}, {Leroux}, {Liao}, {Licata}, {Lindstr{\o}m}, {Lister}, {Livanou},
  {Lobel}, {L{\'o}pez}, {Managau}, {Mann}, {Mantelet}, {Marchal}, {Marchant},
  {Marconi}, {Marinoni}, {Marschalk{\'o}}, {Marshall}, {Martino}, {Marton},
  {Mary}, {Massari}, {Matijevi{\v{c}}}, {Mazeh}, {McMillan}, {Messina},
  {Michalik}, {Millar}, {Molina}, {Molinaro}, {Moln{\'a}r}, {Montegriffo},
  {Mor}, {Morbidelli}, {Morel}, {Morris}, {Mulone}, {Muraveva}, {Musella},
  {Nelemans}, {Nicastro}, {Noval}, {O'Mullane}, {Ord{\'e}novic},
  {Ord{\'o}{\~n}ez-Blanco}, {Osborne}, {Pagani}, {Pagano}, {Pailler},
  {Palacin}, {Palaversa}, {Panahi}, {Pawlak}, {Piersimoni}, {Pineau}, {Plachy},
  {Plum}, {Poggio}, {Poujoulet}, {Pr{\v{s}}a}, {Pulone}, {Racero}, {Ragaini},
  {Rambaux}, {Ramos-Lerate}, {Regibo}, {Reyl{\'e}}, {Riclet}, {Ripepi}, {Riva},
  {Rivard}, {Rixon}, {Roegiers}, {Roelens}, {Romero-G{\'o}mez}, {Rowell},
  {Royer}, {Ruiz-Dern}, {Sadowski}, {Sagrist{\`a} Sell{\'e}s}, {Sahlmann},
  {Salgado}, {Salguero}, {Sanna}, {Santana-Ros}, {Sarasso}, {Savietto},
  {Schultheis}, {Sciacca}, {Segol}, {Segovia}, {S{\'e}gransan}, {Shih},
  {Siltala}, {Silva}, {Smart}, {Smith}, {Solano}, {Solitro}, {Sordo}, {Soria
  Nieto}, {Souchay}, {Spagna}, {Spoto}, {Stampa}, {Steele},
  {Steidelm{\"u}ller}, {Stephenson}, {Stoev}, {Suess}, {Surdej}, {Szabados},
  {Szegedi-Elek}, {Tapiador}, {Taris}, {Tauran}, {Taylor}, {Teixeira},
  {Terrett}, {Teyssand ier}, {Thuillot}, {Titarenko}, {Torra Clotet}, {Turon},
  {Ulla}, {Utrilla}, {Uzzi}, {Vaillant}, {Valentini}, {Valette}, {van Elteren},
  {Van Hemelryck}, {van Leeuwen}, {Vaschetto}, {Vecchiato}, {Veljanoski},
  {Viala}, {Vicente}, {Vogt}, {von Essen}, {Voss}, {Votruba}, {Voutsinas},
  {Walmsley}, {Weiler}, {Wertz}, {Wevers}, {Wyrzykowski}, {Yoldas},
  {{\v{Z}}erjal}, {Ziaeepour}, {Zorec}, {Zschocke}, {Zucker}, {Zurbach}, \&
  {Zwitter}}]{Gaia18_DR2_catalog}
{Gaia Collaboration}, {Brown}, A.~G.~A., {Vallenari}, A., {et~al.} 2018, \aap,
  616, A1

\bibitem[{{Gao} {et~al.}(2014){Gao}, {Liu}, {Zhang}, {Justham}, {Deng}, \&
  {Yang}}]{2014ApJ...788L..37G}
{Gao}, S., {Liu}, C., {Zhang}, X., {et~al.} 2014, \apjl, 788, L37

\bibitem[{{Garraffo} {et~al.}(2018){Garraffo}, {Drake}, {Dotter}, {Choi},
  {Burke}, {Moschou}, {Alvarado-G{\'o}mez}, {Kashyap}, \&
  {Cohen}}]{2018ApJ...862...90G}
{Garraffo}, C., {Drake}, J.~J., {Dotter}, A., {et~al.} 2018, \apj, 862, 90

\bibitem[{{Gary}(2001)}]{gary2001}
{Gary}, G.~A. 2001, \solphys, 203, 71

\bibitem[{{G{\'a}sp{\'a}r} {et~al.}(2016){G{\'a}sp{\'a}r}, {Rieke}, \&
  {Ballering}}]{2016ApJ...826..171G}
{G{\'a}sp{\'a}r}, A., {Rieke}, G.~H., \& {Ballering}, N. 2016, \apj, 826, 171

\bibitem[{{Glebocki} {et~al.}(1980){Glebocki}, {Musielak}, \&
  {Stawikowski}}]{1980AcA....30..453G}
{Glebocki}, R., {Musielak}, G., \& {Stawikowski}, A. 1980, \actaa, 30, 453

\bibitem[{{Gray}(1992)}]{1992PASP..104.1035G}
{Gray}, D.~F. 1992, \pasp, 104, 1035

\bibitem[{{Gray} {et~al.}(2003){Gray}, {Corbally}, {Garrison}, {McFadden}, \&
  {Robinson}}]{Gray03}
{Gray}, R.~O., {Corbally}, C.~J., {Garrison}, R.~F., {McFadden}, M.~T., \&
  {Robinson}, P.~E. 2003, \aj, 126, 2048

\bibitem[{{Grether} \& {Lineweaver}(2006)}]{Grether06_Ref1}
{Grether}, D. \& {Lineweaver}, C.~H. 2006, \apj, 640, 1051

\bibitem[{{Gu} {et~al.}(1997){Gu}, {Jefferies}, {Lindsey}, \&
  {Avrett}}]{1997ApJ...484..960G}
{Gu}, Y., {Jefferies}, J.~T., {Lindsey}, C., \& {Avrett}, E.~H. 1997, \apj,
  484, 960

\bibitem[{{G{\"u}del}(2002)}]{2002ARA&A..40..217G}
{G{\"u}del}, M. 2002, \araa, 40, 217

\bibitem[{{Guedel}(2006)}]{2006astro.ph..9389G}
{Guedel}, M. 2006, arXiv e-prints [\eprint{astro-ph/0609389}]

\bibitem[{{G{\"u}nther} {et~al.}(2020){G{\"u}nther}, {Zhan}, {Seager},
  {Rimmer}, {Ranjan}, {Stassun}, {Oelkers}, {Daylan}, {Newton}, {Kristiansen},
  {Olah}, {Gillen}, {Rappaport}, {Ricker}, {Vanderspek}, {Latham}, {Winn},
  {Jenkins}, {Glidden}, {Fausnaugh}, {Levine}, {Dittmann}, {Quinn},
  {Krishnamurthy}, \& {Ting}}]{Gunther20_Ref19}
{G{\"u}nther}, M.~N., {Zhan}, Z., {Seager}, S., {et~al.} 2020, \aj, 159, 60

\bibitem[{{Hauschildt} \& {Baron}(1999)}]{1999JCoAM.109...41H}
{Hauschildt}, P.~H. \& {Baron}, E. 1999, Journal of Computational and Applied
  Mathematics, 109, 41

\bibitem[{{Henry} {et~al.}(1996){Henry}, {Soderblom}, {Donahue}, \&
  {Baliunas}}]{1996AJ....111..439H}
{Henry}, T.~J., {Soderblom}, D.~R., {Donahue}, R.~A., \& {Baliunas}, S.~L.
  1996, \aj, 111, 439

\bibitem[{{Herrero} {et~al.}(2012){Herrero}, {Ribas}, {Jordi}, {Guinan}, \&
  {Engle}}]{2012A&A...537A.147H}
{Herrero}, E., {Ribas}, I., {Jordi}, C., {Guinan}, E.~F., \& {Engle}, S.~G.
  2012, \aap, 537, A147

\bibitem[{{Hinkel} {et~al.}(2014){Hinkel}, {Timmes}, {Young}, {Pagano}, \&
  {Turnbull}}]{2014AJ....148...54H}
{Hinkel}, N.~R., {Timmes}, F.~X., {Young}, P.~A., {Pagano}, M.~D., \&
  {Turnbull}, M.~C. 2014, \aj, 148, 54

\bibitem[{{H{\o}g} {et~al.}(2000){H{\o}g}, {Fabricius}, {Makarov}, {Urban},
  {Corbin}, {Wycoff}, {Bastian}, {Schwekendiek}, \&
  {Wicenec}}]{2000A&A...355L..27H}
{H{\o}g}, E., {Fabricius}, C., {Makarov}, V.~V., {et~al.} 2000, \aap, 355, L27

\bibitem[{{Holmberg} {et~al.}(2009){Holmberg}, {Nordstr{\"o}m}, \&
  {Andersen}}]{2009A&A...501..941H}
{Holmberg}, J., {Nordstr{\"o}m}, B., \& {Andersen}, J. 2009, \aap, 501, 941

\bibitem[{{Husser} {et~al.}(2013){Husser}, {Wende-von Berg}, {Dreizler},
  {Homeier}, {Reiners}, {Barman}, \& {Hauschildt}}]{2013A&A...553A...6H}
{Husser}, T.-O., {Wende-von Berg}, S., {Dreizler}, S., {et~al.} 2013, \aap,
  553, A6

\bibitem[{{Jao} {et~al.}(2003){Jao}, {Henry}, {Subasavage}, {Bean}, {Costa},
  {Ianna}, \& {M{\'e}ndez}}]{2003AJ....125..332J}
{Jao}, W.-C., {Henry}, T.~J., {Subasavage}, J.~P., {et~al.} 2003, \aj, 125, 332

\bibitem[{{Jao} {et~al.}(2014){Jao}, {Henry}, {Subasavage}, {Winters}, {Gies},
  {Riedel}, \& {Ianna}}]{2014AJ....147...21J}
{Jao}, W.-C., {Henry}, T.~J., {Subasavage}, J.~P., {et~al.} 2014, \aj, 147, 21

\bibitem[{{Kervella} {et~al.}(2017){Kervella}, {Th{\'e}venin}, \&
  {Lovis}}]{2017A&A...598L...7K}
{Kervella}, P., {Th{\'e}venin}, F., \& {Lovis}, C. 2017, \aap, 598, L7

\bibitem[{{Kim} \& {Demarque}(1996)}]{1996ApJ...457..340K}
{Kim}, Y.-C. \& {Demarque}, P. 1996, \apj, 457, 340

\bibitem[{{Klein} {et~al.}(2021){Klein}, {Donati}, {H{\'e}brard}, {Zaire},
  {Folsom}, {Morin}, {Delfosse}, \& {Bonfils}}]{2021MNRAS.500.1844K}
{Klein}, B., {Donati}, J.-F., {H{\'e}brard}, {\'E}.~M., {et~al.} 2021, \mnras,
  500, 1844

\bibitem[{{Lacy} {et~al.}(1976){Lacy}, {Moffett}, \&
  {Evans}}]{Lacy76_Flare_stats_Mstars}
{Lacy}, C.~H., {Moffett}, T.~J., \& {Evans}, D.~S. 1976, \apjs, 30, 85

\bibitem[{{Lawler} {et~al.}(2009){Lawler}, {Beichman}, {Bryden}, {Ciardi},
  {Tanner}, {Su}, {Stapelfeldt}, {Lisse}, \& {Harker}}]{2009ApJ...705...89L}
{Lawler}, S.~M., {Beichman}, C.~A., {Bryden}, G., {et~al.} 2009, \apj, 705, 89

\bibitem[{{Lestrade} \& {Thilliez}(2015)}]{2015A&A...576A..72L}
{Lestrade}, J.-F. \& {Thilliez}, E. 2015, \aap, 576, A72

\bibitem[{{Liebert} {et~al.}(2005){Liebert}, {Young}, {Arnett}, {Holberg}, \&
  {Williams}}]{Liebert05}
{Liebert}, J., {Young}, P.~A., {Arnett}, D., {Holberg}, J.~B., \& {Williams},
  K.~A. 2005, \apjl, 630, L69

\bibitem[{{Lim} {et~al.}(1996){Lim}, {White}, \& {Slee}}]{Lim96_Proxima_ATCA}
{Lim}, J., {White}, S.~M., \& {Slee}, O.~B. 1996, \apj, 460, 976

\bibitem[{{Liseau}(2019)}]{2019arXiv190403043L}
{Liseau}, R. 2019, arXiv e-prints, arXiv:1904.03043

\bibitem[{{Liseau} {et~al.}(2016){Liseau}, {De la Luz}, {O'Gorman}, {Bertone},
  {Chavez}, \& {Tapia}}]{2016A&A...594A.109L}
{Liseau}, R., {De la Luz}, V., {O'Gorman}, E., {et~al.} 2016, \aap, 594, A109

\bibitem[{{Liseau} {et~al.}(2013){Liseau}, {Montesinos}, {Olofsson}, {Bryden},
  {Marshall}, {Ardila}, {Bayo Aran}, {Danchi}, {del Burgo}, {Eiroa}, {Ertel},
  {Fridlund}, {Krivov}, {Pilbratt}, {Roberge}, {Th{\'e}bault}, {Wiegert}, \&
  {White}}]{2013A&A...549L...7L}
{Liseau}, R., {Montesinos}, B., {Olofsson}, G., {et~al.} 2013, \aap, 549, L7

\bibitem[{{Liseau} {et~al.}(2015){Liseau}, {Vlemmings}, {Bayo}, {Bertone},
  {Black}, {del Burgo}, {Chavez}, {Danchi}, {De la Luz}, {Eiroa}, {Ertel},
  {Fridlund}, {Justtanont}, {Krivov}, {Marshall}, {Mora}, {Montesinos},
  {Nyman}, {Olofsson}, {Sanz-Forcada}, {Th{\'e}bault}, \&
  {White}}]{2015A&A...573L...4L}
{Liseau}, R., {Vlemmings}, W., {Bayo}, A., {et~al.} 2015, \aap, 573, L4

\bibitem[{{Lisse} {et~al.}(2007){Lisse}, {Beichman}, {Bryden}, \&
  {Wyatt}}]{2007ApJ...658..584L}
{Lisse}, C.~M., {Beichman}, C.~A., {Bryden}, G., \& {Wyatt}, M.~C. 2007, \apj,
  658, 584

\bibitem[{{Loukitcheva} {et~al.}(2015){Loukitcheva}, {Solanki}, {Carlsson}, \&
  {White}}]{Loukitcheva15_mmRad_model}
{Loukitcheva}, M., {Solanki}, S.~K., {Carlsson}, M., \& {White}, S.~M. 2015,
  \aap, 575, A15

\bibitem[{{Luck}(2017)}]{Luck17}
{Luck}, R.~E. 2017, \aj, 153, 21

\bibitem[{{Lund} {et~al.}(2020){Lund}, {Jardine}, {Lehmann}, {Mackay}, {See},
  {Vidotto}, {Donati}, {Fares}, {Folsom}, {Jeffers}, {Marsden}, {Morin}, \&
  {Petit}}]{Lund20_stelar_helicity}
{Lund}, K., {Jardine}, M., {Lehmann}, L.~T., {et~al.} 2020, \mnras, 493, 1003

\bibitem[{{MacGregor} {et~al.}(2016){MacGregor}, {Lawler}, {Wilner},
  {Matthews}, {Kennedy}, {Booth}, \& {Di Francesco}}]{2016ApJ...828..113M}
{MacGregor}, M.~A., {Lawler}, S.~M., {Wilner}, D.~J., {et~al.} 2016, \apj, 828,
  113

\bibitem[{{MacGregor} {et~al.}(2017){MacGregor}, {Matr{\`a}}, {Kalas},
  {Wilner}, {Pan}, {Kennedy}, {Wyatt}, {Duchene}, {Hughes}, {Rieke}, {Clampin},
  {Fitzgerald}, {Graham}, {Holland}, {Pani{\'c}}, {Shannon}, \&
  {Su}}]{2017ApJ...842....8M}
{MacGregor}, M.~A., {Matr{\`a}}, L., {Kalas}, P., {et~al.} 2017, \apj, 842, 8

\bibitem[{{MacGregor} {et~al.}(2021){MacGregor}, {Weinberger}, {Loyd},
  {Shkolnik}, {Barclay}, {Howard}, {Zic}, {Osten}, {Cranmer}, {Kowalski},
  {Lenc}, {Youngblood}, {Estes}, {Wilner}, {Forbrich}, {Hughes}, {Law},
  {Murphy}, {Boley}, \& {Matthews}}]{2021ApJ...911L..25M}
{MacGregor}, M.~A., {Weinberger}, A.~J., {Loyd}, R.~O.~P., {et~al.} 2021,
  \apjl, 911, L25

\bibitem[{{MacGregor} {et~al.}(2018){MacGregor}, {Weinberger}, {Wilner},
  {Kowalski}, \& {Cranmer}}]{MacGregor18_proxima_flares}
{MacGregor}, M.~A., {Weinberger}, A.~J., {Wilner}, D.~J., {Kowalski}, A.~F., \&
  {Cranmer}, S.~R. 2018, \apjl, 855, L2

\bibitem[{{MacGregor} {et~al.}(2015){MacGregor}, {Wilner}, {Andrews},
  {Lestrade}, \& {Maddison}}]{2015ApJ...809...47M}
{MacGregor}, M.~A., {Wilner}, D.~J., {Andrews}, S.~M., {Lestrade}, J.-F., \&
  {Maddison}, S. 2015, \apj, 809, 47

\bibitem[{{Magaudda} {et~al.}(2020){Magaudda}, {Stelzer}, {Covey}, {Raetz},
  {Matt}, \& {Scholz}}]{Magaudda20_XrayActivity_dMs}
{Magaudda}, E., {Stelzer}, B., {Covey}, K.~R., {et~al.} 2020, \aap, 638, A20

\bibitem[{{Malo} {et~al.}(2014){Malo}, {Artigau}, {Doyon}, {Lafreni{\`e}re},
  {Albert}, \& {Gagn{\'e}}}]{Malo14_Ref7}
{Malo}, L., {Artigau}, {\'E}., {Doyon}, R., {et~al.} 2014, \apj, 788, 81

\bibitem[{{Mamajek}(2012)}]{Mamajek12}
{Mamajek}, E.~E. 2012, \apjl, 754, L20

\bibitem[{Mamajek {et~al.}(2013)Mamajek, Bartlett, Seifahrt, Henry, Dieterich,
  Lurie, Kenworthy, Jao, Riedel, Subasavage, Winters, Finch, Ianna, \&
  Bean}]{Mamajek13_fomalhaut_3rd_companionDiscovery}
Mamajek, E.~E., Bartlett, J.~L., Seifahrt, A., {et~al.} 2013, \apj, 146, 154

\bibitem[{{Mamajek} \& {Hillenbrand}(2008)}]{Mamajek08_Ref11}
{Mamajek}, E.~E. \& {Hillenbrand}, L.~A. 2008, \apj, 687, 1264

\bibitem[{Marino {et~al.}(2017)Marino, Wyatt, Kennedy, Holland, Matrà,
  Shannon, \& Ivison}]{2017MNRAS.469.3518M}
Marino, S., Wyatt, M.~C., Kennedy, G.~M., {et~al.} 2017, Monthly Notices of the
  Royal Astronomical Society, 469, 3518–3531

\bibitem[{{Marino} {et~al.}(2017){Marino}, {Wyatt}, {Pani{\'c}}, {Matr{\`a}},
  {Kennedy}, {Bonsor}, {Kral}, {Dent}, {Duchene}, {Wilner}, {Lisse},
  {Lestrade}, \& {Matthews}}]{2017MNRAS.465.2595M}
{Marino}, S., {Wyatt}, M.~C., {Pani{\'c}}, O., {et~al.} 2017, \mnras, 465, 2595

\bibitem[{{Marsden} {et~al.}(2014){Marsden}, {Petit}, {Jeffers}, {Morin},
  {Fares}, {Reiners}, {do Nascimento}, {Auri{\`e}re}, {Bouvier}, {Carter},
  {Catala}, {Dintrans}, {Donati}, {Gastine}, {Jardine}, {Konstantinova-Antova},
  {Lanoux}, {Ligni{\`e}res}, {Morgenthaler}, {Ram{\`\i}rez-V{\`e}lez},
  {Th{\'e}ado}, {Van Grootel}, \& {BCool Collaboration}}]{2014MNRAS.444.3517M}
{Marsden}, S.~C., {Petit}, P., {Jeffers}, S.~V., {et~al.} 2014, \mnras, 444,
  3517

\bibitem[{{Matr{\`a}} {et~al.}(2017){Matr{\`a}}, {MacGregor}, {Kalas}, {Wyatt},
  {Kennedy}, {Wilner}, {Duchene}, {Hughes}, {Pan}, {Shannon}, {Clampin},
  {Fitzgerald}, {Graham}, {Holland}, {Pani{\'c}}, \&
  {Su}}]{2017ApJ...842....9M}
{Matr{\`a}}, L., {MacGregor}, M.~A., {Kalas}, P., {et~al.} 2017, \apj, 842, 9

\bibitem[{{McMullin} {et~al.}(2007){McMullin}, {Waters}, {Schiebel}, {Young},
  \& {Golap}}]{casa}
{McMullin}, J.~P., {Waters}, B., {Schiebel}, D., {Young}, W., \& {Golap}, K.
  2007, in Astronomical Society of the Pacific Conference Series, Vol. 376,
  Astronomical Data Analysis Software and Systems XVI, ed. R.~A. {Shaw},
  F.~{Hill}, \& D.~J. {Bell}, 127

\bibitem[{{Mermilliod}(1992)}]{Mermilliod92_Ref5}
{Mermilliod}, J.~C. 1992, Highlights of Astronomy, 9, 725

\bibitem[{{Messina} {et~al.}(2017){Messina}, {Millward}, {Buccino}, {Zhang},
  {Medhi}, {Jofr{\'e}}, {Petrucci}, {Pi}, {Hambsch}, {Kehusmaa}, {Harlingten},
  {Artemenko}, {Curtis}, {Hentunen}, {Malo}, {Mauas}, {Monard}, {Muro Serrano},
  {Naves}, {Santallo}, {Savuskin}, \& {Tan}}]{Messina17_Ref6}
{Messina}, S., {Millward}, M., {Buccino}, A., {et~al.} 2017, \aap, 600, A83

\bibitem[{{Millan-Gabet} {et~al.}(2011){Millan-Gabet}, {Serabyn}, {Mennesson},
  {Traub}, {Barry}, {Danchi}, {Kuchner}, {Stark}, {Ragland}, {Hrynevych},
  {Woillez}, {Stapelfeldt}, {Bryden}, {Colavita}, \&
  {Booth}}]{2011ApJ...734...67M}
{Millan-Gabet}, R., {Serabyn}, E., {Mennesson}, B., {et~al.} 2011, \apj, 734,
  67

\bibitem[{{Moerchen} {et~al.}(2007){Moerchen}, {Telesco}, {Packham}, \&
  {Kehoe}}]{2007ApJ...655L.109M}
{Moerchen}, M.~M., {Telesco}, C.~M., {Packham}, C., \& {Kehoe}, T.~J.~J. 2007,
  \apjl, 655, L109

\bibitem[{{Montesinos} {et~al.}(2016){Montesinos}, {Eiroa}, {Krivov},
  {Marshall}, {Pilbratt}, {Liseau}, {Mora}, {Maldonado}, {Wolf}, {Ertel},
  {Bayo}, {Augereau}, {Heras}, {Fridlund}, {Danchi}, {Solano}, {Kirchschlager},
  {del Burgo}, \& {Montes}}]{Montesinos16}
{Montesinos}, B., {Eiroa}, C., {Krivov}, A.~V., {et~al.} 2016, \aap, 593, A51

\bibitem[{{Mora} {et~al.}(2001){Mora}, {Mer{\'\i}n}, {Solano}, {Montesinos},
  {de Winter}, {Eiroa}, {Ferlet}, {Grady}, {Davies}, {Miranda}, {Oudmaijer},
  {Palacios}, {Quirrenbach}, {Harris}, {Rauer}, {Collier Cameron}, {Deeg},
  {Garz{\'o}n}, {Penny}, {Schneider}, {Tsapras}, \& {Wesselius}}]{Mora01}
{Mora}, A., {Mer{\'\i}n}, B., {Solano}, E., {et~al.} 2001, \aap, 378, 116

\bibitem[{{Morin} {et~al.}(2010){Morin}, {Donati}, {Petit}, {Delfosse},
  {Forveille}, \& {Jardine}}]{2010MNRAS.407.2269M}
{Morin}, J., {Donati}, J.~F., {Petit}, P., {et~al.} 2010, \mnras, 407, 2269

\bibitem[{{Netopil}(2017)}]{2017MNRAS.469.3042N}
{Netopil}, M. 2017, \mnras, 469, 3042

\bibitem[{{Nicolet}(1975)}]{1975A&AS...22..239N}
{Nicolet}, B. 1975, \aaps, 22, 239

\bibitem[{{Nindos}(2020)}]{2020FrASS...7...57N}
{Nindos}, A. 2020, Frontiers in Astronomy and Space Sciences, 7, 57

\bibitem[{Nindos {et~al.}(2000)Nindos, Kundu, White, Shibasaki, \&
  Gopalswamy}]{Nindos00_SXR_gyroresonance_Sunspots}
Nindos, A., Kundu, M.~R., White, S.~M., Shibasaki, K., \& Gopalswamy, N. 2000,
  \apjs, 130, 485

\bibitem[{{Noyes} {et~al.}(1984){Noyes}, {Hartmann}, {Baliunas}, {Duncan}, \&
  {Vaughan}}]{Noyes84_RHK}
{Noyes}, R.~W., {Hartmann}, L.~W., {Baliunas}, S.~L., {Duncan}, D.~K., \&
  {Vaughan}, A.~H. 1984, \apj, 279, 763

\bibitem[{{Osten} {et~al.}(2005){Osten}, {Hawley}, {Allred}, {Johns-Krull}, \&
  {Roark}}]{2005ApJ...621..398O}
{Osten}, R.~A., {Hawley}, S.~L., {Allred}, J.~C., {Johns-Krull}, C.~M., \&
  {Roark}, C. 2005, \apj, 621, 398

\bibitem[{{Pace}(2013)}]{Pace13_RHK_Drastic_variability}
{Pace}, G. 2013, \aap, 551, L8

\bibitem[{{Perley} {et~al.}(2011){Perley}, {Chandler}, {Butler}, \&
  {Wrobel}}]{Perley11_EVLA}
{Perley}, R.~A., {Chandler}, C.~J., {Butler}, B.~J., \& {Wrobel}, J.~M. 2011,
  \apjl, 739, L1

\bibitem[{{Raghavan} {et~al.}(2010){Raghavan}, {McAlister}, {Henry}, {Latham},
  {Marcy}, {Mason}, {Gies}, {White}, \& {ten Brummelaar}}]{Raghavan10_Ref16}
{Raghavan}, D., {McAlister}, H.~A., {Henry}, T.~J., {et~al.} 2010, \apjs, 190,
  1

\bibitem[{{Ram{\'\i}rez} {et~al.}(2013){Ram{\'\i}rez}, {Allende Prieto}, \&
  {Lambert}}]{2013ApJ...764...78R}
{Ram{\'\i}rez}, I., {Allende Prieto}, C., \& {Lambert}, D.~L. 2013, \apj, 764,
  78

\bibitem[{{Ram{\'\i}rez} {et~al.}(2012){Ram{\'\i}rez}, {Fish}, {Lambert}, \&
  {Allende Prieto}}]{Ramirez12_Ref15}
{Ram{\'\i}rez}, I., {Fish}, J.~R., {Lambert}, D.~L., \& {Allende Prieto}, C.
  2012, \apj, 756, 46

\bibitem[{{Riaz} {et~al.}(2006){Riaz}, {Gizis}, \& {Harvin}}]{Riaz06_Ref9}
{Riaz}, B., {Gizis}, J.~E., \& {Harvin}, J. 2006, \aj, 132, 866

\bibitem[{{Ribas} {et~al.}(2017){Ribas}, {Gregg}, {Boyajian}, \&
  {Bolmont}}]{Ribas17_ProximaCenSED}
{Ribas}, I., {Gregg}, M.~D., {Boyajian}, T.~S., \& {Bolmont}, E. 2017, \aap,
  603, A58

\bibitem[{{Riedel} {et~al.}(2017){Riedel}, {Alam}, {Rice}, {Cruz}, \&
  {Henry}}]{2017ApJ...840...87R}
{Riedel}, A.~R., {Alam}, M.~K., {Rice}, E.~L., {Cruz}, K.~L., \& {Henry}, T.~J.
  2017, \apj, 840, 87

\bibitem[{{Riedel} {et~al.}(2014){Riedel}, {Finch}, {Henry}, {Subasavage},
  {Jao}, {Malo}, {Rodriguez}, {White}, {Gies}, {Dieterich}, {Winters},
  {Davison}, {Nelan}, {Blunt}, {Cruz}, {Rice}, \&
  {Ianna}}]{2014AJ....147...85R}
{Riedel}, A.~R., {Finch}, C.~T., {Henry}, T.~J., {et~al.} 2014, \aj, 147, 85

\bibitem[{{Rodr{\'\i}guez} {et~al.}(2019){Rodr{\'\i}guez}, {Lizano}, {Loinard},
  {Ch{\'a}vez-Dagostino}, {Bastian}, \& {Beasley}}]{2019ApJ...871..172R}
{Rodr{\'\i}guez}, L.~F., {Lizano}, S., {Loinard}, L., {et~al.} 2019, \apj, 871,
  172

\bibitem[{{Royer} {et~al.}(2002){Royer}, {Grenier}, {Baylac}, {G{\'o}mez}, \&
  {Zorec}}]{Royer02}
{Royer}, F., {Grenier}, S., {Baylac}, M.~O., {G{\'o}mez}, A.~E., \& {Zorec}, J.
  2002, \aap, 393, 897

\bibitem[{{Schmitt} \& {Liefke}(2004)}]{Schmitt04_Ref13}
{Schmitt}, J.~H.~M.~M. \& {Liefke}, C. 2004, \aap, 417, 651

\bibitem[{{Schr{\"o}der} \& {Schmitt}(2007)}]{2007A&A...475..677S}
{Schr{\"o}der}, C. \& {Schmitt}, J.~H.~M.~M. 2007, \aap, 475, 677

\bibitem[{Selhorst {et~al.}(2014)Selhorst, Costa, de~Castro, Valio, Pacini, \&
  Shibasaki}]{Selhorst14_NoRH_SSN_3GHz_srcCount}
Selhorst, C.~L., Costa, J. E.~R., de~Castro, C. G.~G., {et~al.} 2014, The
  Astrophysical Journal, 790, 134

\bibitem[{{Selhorst} {et~al.}(2008){Selhorst}, {Silva-V{\'a}lio}, \&
  {Costa}}]{2008A&A...488.1079S}
{Selhorst}, C.~L., {Silva-V{\'a}lio}, A., \& {Costa}, J.~E.~R. 2008, \aap, 488,
  1079

\bibitem[{{Shibasaki} {et~al.}(2011){Shibasaki}, {Alissandrakis}, \&
  {Pohjolainen}}]{2011SoPh..273..309S}
{Shibasaki}, K., {Alissandrakis}, C.~E., \& {Pohjolainen}, S. 2011, \solphys,
  273, 309

\bibitem[{{Shibasaki} {et~al.}(1994){Shibasaki}, {Takano}, {Enome}, {Nakajima},
  {Nishio}, {Hanaoka}, {Torii}, {Sekiguchi}, {Bushimata}, {Kawashima},
  {Shinohara}, {Koshiishi}, \& {Shiomi}}]{1994SSRv...68..217S}
{Shibasaki}, K., {Takano}, T., {Enome}, S., {et~al.} 1994, \ssr, 68, 217

\bibitem[{{S{\"o}derhjelm}(1999)}]{1999A&A...341..121S}
{S{\"o}derhjelm}, S. 1999, \aap, 341, 121

\bibitem[{{Soubiran} {et~al.}(2016){Soubiran}, {Le Campion}, {Brouillet}, \&
  {Chemin}}]{Soubiran16_Ref3}
{Soubiran}, C., {Le Campion}, J.-F., {Brouillet}, N., \& {Chemin}, L. 2016,
  \aap, 591, A118

\bibitem[{{Sreejith} {et~al.}(2020){Sreejith}, {Fossati}, {Youngblood},
  {France}, \& {Ambily}}]{Sreejith20_RHK_indices}
{Sreejith}, A.~G., {Fossati}, L., {Youngblood}, A., {France}, K., \& {Ambily},
  S. 2020, \aap, 644, A67

\bibitem[{{Stassun} {et~al.}(2019){Stassun}, {Oelkers}, {Paegert}, {Torres},
  {Pepper}, {De Lee}, {Collins}, {Latham}, {Muirhead}, {Chittidi},
  {Rojas-Ayala}, {Fleming}, {Rose}, {Tenenbaum}, {Ting}, {Kane}, {Barclay},
  {Bean}, {Brassuer}, {Charbonneau}, {Ge}, {Lissauer}, {Mann}, {McLean},
  {Mullally}, {Narita}, {Plavchan}, {Ricker}, {Sasselov}, {Seager}, {Sharma},
  {Shiao}, {Sozzetti}, {Stello}, {Vanderspek}, {Wallace}, \&
  {Winn}}]{Satssun19_Ref12}
{Stassun}, K.~G., {Oelkers}, R.~J., {Paegert}, M., {et~al.} 2019, \aj, 158, 138

\bibitem[{{Stepien}(1994)}]{stepien94_Defn_Rx+Ro_Vs_activity_n_manyCorCurves}
{Stepien}, K. 1994, \aap, 292, 191

\bibitem[{{Su} {et~al.}(2016){Su}, {Rieke}, {Defr{\'e}re}, {Wang}, {Lai},
  {Wilner}, {van Lieshout}, \& {Lee}}]{2016ApJ...818...45S}
{Su}, K. Y.~L., {Rieke}, G.~H., {Defr{\'e}re}, D., {et~al.} 2016, \apj, 818, 45

\bibitem[{{Suchkov} {et~al.}(2003){Suchkov}, {Makarov}, \&
  {Voges}}]{Suchkov03_Ref18}
{Suchkov}, A.~A., {Makarov}, V.~V., \& {Voges}, W. 2003, \apj, 595, 1206

\bibitem[{{Suresh} {et~al.}(2020){Suresh}, {Chatterjee}, {Cordes}, {Bastian},
  \& {Hallinan}}]{suresh20_EpsEri_RadioSEDmodel}
{Suresh}, A., {Chatterjee}, S., {Cordes}, J.~M., {Bastian}, T.~S., \&
  {Hallinan}, G. 2020, \apj, 904, 138

\bibitem[{{Takeda} {et~al.}(2007){Takeda}, {Ford}, {Sills}, {Rasio}, {Fischer},
  \& {Valenti}}]{Takeda07_Ref17}
{Takeda}, G., {Ford}, E.~B., {Sills}, A., {et~al.} 2007, \apjs, 168, 297

\bibitem[{{Testa}(2010)}]{2010PNAS..107.7158T}
{Testa}, P. 2010, Proceedings of the National Academy of Science, 107, 7158

\bibitem[{Thekaekara {et~al.}(1969)Thekaekara, Kruger, \&
  Duncan}]{Thekaekara69}
Thekaekara, M.~P., Kruger, R., \& Duncan, C.~H. 1969, Appl. Opt., 8, 1713

\bibitem[{{Th{\'e}venin} {et~al.}(2002){Th{\'e}venin}, {Provost}, {Morel},
  {Berthomieu}, {Bouchy}, \& {Carrier}}]{2002A&A...392L...9T}
{Th{\'e}venin}, F., {Provost}, J., {Morel}, P., {et~al.} 2002, \aap, 392, L9

\bibitem[{{Tian} {et~al.}(2020){Tian}, {El-Badry}, {Rix}, \&
  {Gould}}]{2020ApJS..246....4T}
{Tian}, H.-J., {El-Badry}, K., {Rix}, H.-W., \& {Gould}, A. 2020, \apjs, 246, 4

\bibitem[{{Torres} {et~al.}(2006){Torres}, {Quast}, {da Silva}, {de La Reza},
  {Melo}, \& {Sterzik}}]{2006A&A...460..695T}
{Torres}, C.~A.~O., {Quast}, G.~R., {da Silva}, L., {et~al.} 2006, \aap, 460,
  695

\bibitem[{{Trigilio} {et~al.}(2018){Trigilio}, {Umana}, {Cavallaro},
  {Agliozzo}, {Leto}, {Buemi}, {Ingallinera}, {Bufano}, \&
  {Riggi}}]{2018MNRAS.481..217T}
{Trigilio}, C., {Umana}, G., {Cavallaro}, F., {et~al.} 2018, \mnras, 481, 217

\bibitem[{{Tu} {et~al.}(2015){Tu}, {Johnstone}, {G{\"u}del}, \&
  {Lammer}}]{2015A&A...577L...3T}
{Tu}, L., {Johnstone}, C.~P., {G{\"u}del}, M., \& {Lammer}, H. 2015, \aap, 577,
  L3

\bibitem[{{Ujjwal} {et~al.}(2020){Ujjwal}, {Kartha}, {Mathew}, {Manoj}, \&
  {Narang}}]{Ujjwal20}
{Ujjwal}, K., {Kartha}, S.~S., {Mathew}, B., {Manoj}, P., \& {Narang}, M. 2020,
  \aj, 159, 166

\bibitem[{{van Leeuwen}(2007)}]{2007A&A...474..653V}
{van Leeuwen}, F. 2007, \aap, 474, 653

\bibitem[{{Vernazza} {et~al.}(1981){Vernazza}, {Avrett}, \&
  {Loeser}}]{1981ApJS...45..635V}
{Vernazza}, J.~E., {Avrett}, E.~H., \& {Loeser}, R. 1981, \apjs, 45, 635

\bibitem[{{Vican}(2012)}]{2012AJ....143..135V}
{Vican}, L. 2012, \aj, 143, 135

\bibitem[{{Vidotto} {et~al.}(2014){Vidotto}, {Gregory}, {Jardine}, {Donati},
  {Petit}, {Morin}, {Folsom}, {Bouvier}, {Cameron}, {Hussain}, {Marsden},
  {Waite}, {Fares}, {Jeffers}, \& {do Nascimento}}]{Vidotto14_B_Vs_age_n_rot}
{Vidotto}, A.~A., {Gregory}, S.~G., {Jardine}, M., {et~al.} 2014, \mnras, 441,
  2361

\bibitem[{{Villadsen} \& {Hallinan}(2019)}]{2019ApJ...871..214V}
{Villadsen}, J. \& {Hallinan}, G. 2019, \apj, 871, 214

\bibitem[{{Villadsen} {et~al.}(2014){Villadsen}, {Hallinan}, {Bourke},
  {G{\"u}del}, \& {Rupen}}]{Villadsen14_First_detect_SLS_inRadio_VLA}
{Villadsen}, J., {Hallinan}, G., {Bourke}, S., {G{\"u}del}, M., \& {Rupen}, M.
  2014, \apj, 788, 112

\bibitem[{{Vogt} {et~al.}(2010){Vogt}, {Wittenmyer}, {Butler}, {O'Toole},
  {Henry}, {Rivera}, {Meschiari}, {Laughlin}, {Tinney}, {Jones}, {Bailey},
  {Carter}, \& {Batygin}}]{Vogt10}
{Vogt}, S.~S., {Wittenmyer}, R.~A., {Butler}, R.~P., {et~al.} 2010, \apj, 708,
  1366

\bibitem[{{Vourlidas} {et~al.}(2006){Vourlidas}, {Gary}, \&
  {Shibasaki}}]{2006PASJ...58...11V}
{Vourlidas}, A., {Gary}, D.~E., \& {Shibasaki}, K. 2006, \pasj, 58, 11

\bibitem[{{Wedemeyer} {et~al.}(2016){Wedemeyer}, {Bastian}, {Braj{\v{s}}a},
  {Hudson}, {Fleishman}, {Loukitcheva}, {Fleck}, {Kontar}, {De Pontieu},
  {Yagoubov}, {Tiwari}, {Soler}, {Black}, {Antolin}, {Scullion}, {Gun{\'a}r},
  {Labrosse}, {Ludwig}, {Benz}, {White}, {Hauschildt}, {Doyle}, {Nakariakov},
  {Ayres}, {Heinzel}, {Karlicky}, {Van Doorsselaere}, {Gary}, {Alissandrakis},
  {Nindos}, {Solanki}, {Rouppe van der Voort}, {Shimojo}, {Kato},
  {Zaqarashvili}, {Perez}, {Selhorst}, \& {Barta}}]{2016SSRv..200....1W}
{Wedemeyer}, S., {Bastian}, T., {Braj{\v{s}}a}, R., {et~al.} 2016, \ssr, 200, 1

\bibitem[{{White} {et~al.}(2019){White}, {Aufdenberg}, {Boley}, {Devlin},
  {Dicker}, {Hauschildt}, {Hughes}, {Hughes}, {Mason}, {Matthews}, {Mo{\'o}r},
  {Mroczkowski}, {Romero}, {Sievers}, {Stanchfield}, {Tapia}, \&
  {Wilner}}]{2019ApJ...875...55W}
{White}, J.~A., {Aufdenberg}, J., {Boley}, A.~C., {et~al.} 2019, \apj, 875, 55

\bibitem[{{White} {et~al.}(2018){White}, {Aufdenberg}, {Boley}, {Hauschildt},
  {Hughes}, {Matthews}, \& {Wilner}}]{2018ApJ...859..102W}
{White}, J.~A., {Aufdenberg}, J., {Boley}, A.~C., {et~al.} 2018, \apj, 859, 102

\bibitem[{{White} {et~al.}(2017{\natexlab{a}}){White}, {Boley}, {Dent}, {Ford},
  \& {Corder}}]{2017MNRAS.466.4201W}
{White}, J.~A., {Boley}, A.~C., {Dent}, W.~R.~F., {Ford}, E.~B., \& {Corder},
  S. 2017{\natexlab{a}}, \mnras, 466, 4201

\bibitem[{{White} {et~al.}(2020){White}, {Tapia-V{\'a}zquez}, {Hughes},
  {Mo{\'o}r}, {Matthews}, {Wilner}, {Aufdenberg}, {Hughes}, {De la Luz}, \&
  {Boley}}]{White20_MESAS}
{White}, J.~A., {Tapia-V{\'a}zquez}, F., {Hughes}, A.~G., {et~al.} 2020, \apj,
  894, 76

\bibitem[{{White}(2004)}]{2004NewAR..48.1319W}
{White}, S.~M. 2004, \nar, 48, 1319

\bibitem[{{White} {et~al.}(2017{\natexlab{b}}){White}, {Iwai}, {Phillips},
  {Hills}, {Hirota}, {Yagoubov}, {Siringo}, {Shimojo}, {Bastian}, {Hales},
  {Sawada}, {Asayama}, {Sugimoto}, {Marson}, {Kawasaki}, {Muller}, {Nakazato},
  {Sugimoto}, {Braj{\v{s}}a}, {Skoki{\'c}}, {B{\'a}rta}, {Kim}, {Remijan}, {de
  Gregorio}, {Corder}, {Hudson}, {Loukitcheva}, {Chen}, {De Pontieu},
  {Fleishmann}, {Gary}, {Kobelski}, {Wedemeyer}, \&
  {Yan}}]{2017SoPh..292...88W}
{White}, S.~M., {Iwai}, K., {Phillips}, N.~M., {et~al.} 2017{\natexlab{b}},
  \solphys, 292, 88

\bibitem[{{White} \& {Kundu}(1997)}]{1997SoPh..174...31W}
{White}, S.~M. \& {Kundu}, M.~R. 1997, \solphys, 174, 31

\bibitem[{{White} {et~al.}(1991){White}, {Kundu}, \&
  {Gopalswamy}}]{1991ApJ...366L..43W}
{White}, S.~M., {Kundu}, M.~R., \& {Gopalswamy}, N. 1991, \apjl, 366, L43

\bibitem[{{White} {et~al.}(2006){White}, {Loukitcheva}, \&
  {Solanki}}]{2006A&A...456..697W}
{White}, S.~M., {Loukitcheva}, M., \& {Solanki}, S.~K. 2006, \aap, 456, 697

\bibitem[{{Wiegert} {et~al.}(2014){Wiegert}, {Liseau}, {Th{\'e}bault},
  {Olofsson}, {Mora}, {Bryden}, {Marshall}, {Eiroa}, {Montesinos}, {Ardila},
  {Augereau}, {Bayo Aran}, {Danchi}, {del Burgo}, {Ertel}, {Fridlund},
  {Hajigholi}, {Krivov}, {Pilbratt}, {Roberge}, {White}, \& {Wolf}}]{Wiegert14}
{Wiegert}, J., {Liseau}, R., {Th{\'e}bault}, P., {et~al.} 2014, \aap, 563, A102

\bibitem[{{Wright} {et~al.}(2004){Wright}, {Marcy}, {Butler}, \&
  {Vogt}}]{Wright04_Ref2}
{Wright}, J.~T., {Marcy}, G.~W., {Butler}, R.~P., \& {Vogt}, S.~S. 2004, \apjs,
  152, 261

\bibitem[{{Wright} {et~al.}(2011){Wright}, {Drake}, {Mamajek}, \&
  {Henry}}]{Wright11_Ref10}
{Wright}, N.~J., {Drake}, J.~J., {Mamajek}, E.~E., \& {Henry}, G.~W. 2011,
  \apj, 743, 48

\bibitem[{{Wyatt} {et~al.}(2012){Wyatt}, {Kennedy}, {Sibthorpe},
  {Moro-Mart{\'\i}n}, {Lestrade}, {Ivison}, {Matthews}, {Udry}, {Greaves},
  {Kalas}, {Lawler}, {Su}, {Rieke}, {Booth}, {Bryden}, {Horner}, {Kavelaars},
  \& {Wilner}}]{Wyatt12}
{Wyatt}, M.~C., {Kennedy}, G., {Sibthorpe}, B., {et~al.} 2012, \mnras, 424,
  1206

\bibitem[{{Zacharias} {et~al.}(2012){Zacharias}, {Finch}, {Girard}, {Henden},
  {Bartlett}, {Monet}, \& {Zacharias}}]{2012yCat.1322....0Z}
{Zacharias}, N., {Finch}, C.~T., {Girard}, T.~M., {et~al.} 2012, VizieR Online
  Data Catalog, I/322A

\bibitem[{{Zic} {et~al.}(2020){Zic}, {Murphy}, {Lynch}, {Heald}, {Lenc},
  {Kaplan}, {Cairns}, {Coward}, {Gendre}, {Johnston}, {MacGregor}, {Price}, \&
  {Wheatland}}]{Zic20_typeIV_ProximaCen}
{Zic}, A., {Murphy}, T., {Lynch}, C., {et~al.} 2020, \apj, 905, 23

\bibitem[{{Zirin} {et~al.}(1991){Zirin}, {Baumert}, \&
  {Hurford}}]{1991ApJ...370..779Z}
{Zirin}, H., {Baumert}, B.~M., \& {Hurford}, G.~J. 1991, \apj, 370, 779

\bibitem[{{Zorec} \& {Royer}(2012)}]{2012A&A...537A.120Z}
{Zorec}, J. \& {Royer}, F. 2012, \aap, 537, A120

\end{thebibliography}

\end{document}